\documentclass[letter,10pt]{article}
\usepackage[top=1.0in, bottom=1.0in, left=1.0in, right=1.0in]{geometry}
\usepackage{times}
\usepackage[hang,flushmargin,stable]{footmisc}
\usepackage{graphicx}
\usepackage{amssymb}
\usepackage{amsmath}
\usepackage{colortbl}
\usepackage{color}
\usepackage[bf]{caption}
\usepackage{times}
\usepackage{indentfirst}
\usepackage{url}
\usepackage{subfigure}
\usepackage{tikz}
\usepackage{bbding}
\usepackage[all]{xy}
\usepackage{float}
\usepackage{array}
\usepackage{paralist}

\usepackage{multirow}

\makeatletter
\def\url@leostyle{%
  \@ifundefined{selectfont}{\def\UrlFont{\small}}%
  {\def\UrlFont{\small}}%
}
\makeatother	
\definecolor{darkblue}{RGB}{0,0,135}
\usepackage[bookmarks=false, colorlinks=true, plainpages = false, linkcolor = darkblue,   citecolor = darkblue, urlcolor = darkblue, filecolor = darkblue]{hyperref}

\newcommand{\A}{\ensuremath{{\sf Alice}}}
\newcommand{\B}{\ensuremath{{\sf Bob}}}
\newcommand{\mmm}{\mbox{\textsf{\em m}}}
\newcommand{\ignore}[1]{}

\newcommand{\descr}[1]{\vspace{0.45cm} \noindent \textbf{#1}}

\newcommand{\spa}{\mbox{ }}
\newcommand{\equal}{\hspace*{-0.05cm}=\hspace*{-0.05cm}}

\newcommand{\DGT}{{\sf DGT12}}

\newcommand{\Z}{\mathbb{Z}}

\title{\bf EsPRESSo: Ef\/ficient Privacy-Preserving Evaluation\\of Sample Set Similarity\thanks{\bf\em A
preliminary version of this paper was published in the Proceedings of the 7th ESORICS International
Workshop on Digital Privacy Management (DPM 2012). This is the full version, appearing in the Journal of Computer Security.}} 
\author{\normalsize Carlo Blundo$^{1}$, Emiliano De Cristofaro$^{2}$, and Paolo Gasti$^3$\vspace{0.15cm}\\
{\normalsize $^1$ Universit\`a di Salerno, Italy
$\;\;^2$ PARC
$\;\;^3$ New York Institute of Technology}}

\begin{document}
\pagenumbering{arabic}
\thispagestyle{plain}
\date{}
\maketitle

Electronic information is increasingly often shared among entities without complete mutual trust. 
To address related security and privacy issues, a few cryptographic techniques have emerged that support
privacy-preserving information sharing and retrieval. One interesting open problem in this context involves two parties that need to assess the {\em similarity} of their datasets, but are reluctant to disclose their actual content.
This paper presents an efficient and provably-secure construction supporting the privacy-preserving evaluation of sample set similarity, where similarity is measured as the {\em Jaccard} index. We present two protocols: the first securely computes the (Jaccard) similarity of two sets, and 
the second approximates it, using MinHash techniques, with lower complexities. 
We show that our novel protocols are attractive in many compelling applications, including document/multimedia similarity, biometric authentication, and genetic tests. In the process, we demonstrate that our constructions are appreciably more efficient than prior work.

\section{Introduction}\label{sec:intro}
The availability of electronic information is essential to the 
functioning of our communities. Increasingly often, data needs to be shared between
parties without complete mutual trust. Naturally, this raises important privacy concerns 
with respect to the disclosure and the long-term safety of sensitive content.
One interesting problem occurs whenever two or more entities
need to evaluate the similarity of their datasets, but are 
reluctant to openly disclose their data.
This task faces three important technical challenges: (1) how to identify a meaningful metric
to estimate similarity, (2) how to compute a measure thereof such that no 
private information is revealed during the process, and (3) how to do so efficiently.
We address such challenges by introducing a cryptographic primitive called {\bf\em EsPRESSo -- Privacy-Preserving Evaluation of Sample Set Similarity}. Among others, this construction is appealing in 
a few relevant applications, presented below. %

\paragraph{Document similarity:} Two parties need to estimate the similarity
of their documents, or collections thereof.
In many settings, documents contain sensitive information and parties may be unwilling, 
or simply forbidden, to reveal their content. For instance, program chairs of a conference may want
to verify that none of submitted papers is also under review in other conferences or journals,
but, obviously, they are not allowed to disclose papers in submission.
Likewise, two law enforcement authorities (e.g., the FBI and local police), or two investigation teams
with different clearance levels, might need to share documents pertaining suspect terrorists,
but they can do so only conditioned upon a clear indication that content is relevant to the same investigation.

\paragraph{Iris Matching:} Biometric identification and authentication are increasingly used
due to fast and inexpensive devices that can extract biometric information from a multitude of sources, e.g., voice, fingerprints, iris, and so on. Clearly, given its utmost sensitivity, biometric data must
be protected from arbitrary disclosure. 
Consider, for instance, an agency that needs to determine whether a given biometric appears on a government watch-list.
As agencies may have different clearance levels, privacy of biometric's owner needs to be preserved if no matches are found, but, at the same time, unrestricted access to the watch-list cannot be granted.

\paragraph{Multimedia File Similarity:} 
Digital media, e.g., images, audio, video,
are increasingly relevant in today's computing ecosystems.
Consider two parties that wish to evaluate similarity of their media files, e.g.,
for plagiarism detection: sensitivity of possibly unreleased material (or copyright issues)
may prevent parties from revealing actual content.

\bigskip

EsPRESSO does not only appeal to examples above, but are also relevant to a wide
spectrum of applications, for instance, in the context of privacy-preserving sharing of information and/or recommender systems, e.g., to privately assess similarity of genomic information~\cite{baldi2011countering}, social network profiles~\cite{pino}, attackers' information~\cite{katti2005collaborating}, etc.

\subsection{Technical Roadmap \& Contributions}
Our first step is to identify a {\em metric} for effectively evaluating
similarity of sample sets. Several similarity measures are available and commonly
used in different contexts,
such as Cosine, Euclidean, Manhattan, Minkowski similarity, or Hamming and Levenshtein distances.
In this paper, we focus on a well-known metric, namely, the \emph{Jaccard Similarity Index}~\cite{jaccard1901etude}, which quantifies the similarity of {\em any} two sets $A$ and $B$. It is expressed as
a rational number between 0 and 1, and, as showed in~\cite{broder1997resemblance}, it effectively captures the informal notion of ``roughly the same''. The Jaccard Index can be used, e.g.,
to find near duplicate records~\cite{xiao2008efficient}
and similar documents~\cite{broder1997resemblance},
for web-page clustering~\cite{strehl2000impact}, 
data mining~\cite{tan2006introduction},
and genetic tests~\cite{genodroid,dombek2000use,popescu2006fuzzy}.
Also note that, as sample sets can be relatively large, in distributed settings
an approximation of the index is oftentimes preferred to its exact calculation.
To this end, \emph{MinHash} techniques~\cite{broder1997resemblance} are often used to estimate
the Jaccard index, with remarkably lower computation and communication costs (see Section~\ref{subsec:minhash}).

We define and instantiate a cryptographic primitive for efficient privacy-preserving evaluation of sample set similarity (or EsPRESSo, for short). We present two instantiations, that
allow two interacting parties to compute and/or approximate the Jaccard similarity of their private sets,
without reciprocally disclosing any information about their content (or, at most, their size).
Our main cryptographic building block is {\em Private Set Intersection
Cardinality} (PSI-CA)~\cite{freedman2004efficient}, which we review in Section~\ref{subsec:psi-ca}.
Specifically, we use PSI-CA to privately compute the magnitude of set intersection and union,
and we then derive the value of the Jaccard index.
As fast (linear-complexity) PSI-CA protocols become available (e.g.,~\cite{ePrint}), %
this can be done efficiently, even on large sets.
Nonetheless, our work shows that, using MinHash approximations,
one can obtain an estimate of the Jaccard index with remarkably increased efficiency,
by reducing the size of input sets (thus, the number of underlying cryptographic operations).

Privacy-preserving evaluation of sample set similarity is appealing in many scenarios.
We focus on document and multimedia similarity as well as iris matching,
and show that privacy is attainable with low overhead.
Experiments demonstrate that our generic technique -- while not bounded to any specific application --
is appreciably more efficient than state-of-the-art protocols that only focus
on one specific scenario, while maintaing comparable accuracy.
Finally, in the process of reviewing related work, we identify limits and flaws of some prior results. 

\paragraph{Organization.} The rest of this paper is organized as follows.
Next section introduces building blocks, then 
Section~\ref{sec:new} presents our construction for secure computation of Jaccard index 
and an even more efficient technique to (privately) approximate it.
Then, Sections~\ref{sec:docsimilarity}, \ref{sec:iris}, and~\ref{sec:multimedia} present
our constructions for privacy-preserving similarity evaluation of, respectively, documents, irises,
and multimedia content. 
Finally, Section~\ref{sec:approx-psica} sketches a very efficient 
protocol that privately approximates set intersection cardinality, additionally hiding input set sizes,
while the paper concludes in Section~\ref{sec:conclusion}. Appendix~\ref{app:minhash} presents some more details
on MinHash, and Appendix~\ref{app:flaw} shows a flaw in the protocol for secure document similarity in~\cite{JiSa}.

\section{Preliminaries}\label{sec:preliminaries}
This section provides some relevant background information on Jaccard index, MinHash techniques, and our main cryptographic building blocks.

\subsection{Jaccard Similarity Index and MinHash Techniques}

\paragraph{Jaccard Index. }\label{subsec:jaccard} 
One of the most common metrics for assessing the similarity of two sets  $A$ and $B$ (hence, of 
data they represent) is the Jaccard index~\cite{jaccard1901etude}, defined as $J(A,B)=|A\cap B|/|A\cup B|$.
Values close to 1 suggest that two sets are 
very similar, whereas, those closer to 0 indicate that $A$ and $B$ are almost disjoint.
Note that the Jaccard index of $A$ and $B$ can be rewritten as a mere function of the {\em intersection}:
$J(A,B)=|A\cap B|/(|A|+|B|-|A\cap B|)$.

\paragraph{MinHash Techniques.}\label{subsec:minhash}
Clearly, computing the Jaccard index incurs a complexity linear in set sizes.
Thus, in the context of a large number of big sets, its computation might be relatively expensive.
In fact, for each pair of sets, the Jaccard index must be computed from scratch,
i.e., no information used to calculate $J(A,B)$ can be re-used for $J(A,C)$. 
(That is, similarity is not a transitive measure.)
As a result, an {\em approximation} of the Jaccard index is often preferred,
as it can be obtained at a significantly lower cost, e.g., using
MinHash~\cite{broder1997resemblance}.
Informally, MinHash techniques extract a small representation $h_{k}(S)$ of a set $S$ through deterministic 
(salted) sampling. This representation has a constant size $k$, independent from $|S|$, and can be used to compute an approximation of the Jaccard index. The parameter $k$ also defines %
the expected error with respect to the exact Jaccard index. Intuitively, larger values of $k$
yield smaller approximation errors.
The computation of $h_{k}(S)$ also incurs a complexity linear in set sizes, however,
it must be performed {\em only once} per set, for {\em any} number of comparisons. 
Thus, with MinHash techniques, evaluating the similarity of any two sets requires only a constant number  of comparisons.
Similarly, the bandwidth used by two interacting parties to approximate the Jaccard index 
of their respective sets is also constant (i.e., $O(k)$).

There are two  strategies to realize MinHashes: one employs multiple hash functions,
while the other is built from a single hash function. This paper focuses on the former 
technique. Thus, we defer the description of the latter to Appendix~\ref{app:minhash}, which 
also overviews possible MinHash instantiations.

\paragraph{MinHash with many hash functions.} 
Let $\mathcal{F}$ be a family of hash functions that map items from set $U$ to distinct $\tau$-bit integers.
Select $k$ different functions $h^{(1)}(\cdot),\ldots, h^{(k)}(\cdot)$ from $\mathcal{F}$;
for any set $S\subseteq U$, let $h^{(i)}_{min}(S)$ be the item $s \in S$ with the smallest value $h^{(i)}(s)$ .
The MinHash representation $h_k(S)$ of set $S$ is a vector $h_k(S) = \{h_{min}^{(i)}(S)\}_{i=1}^k$.
The Jaccard index $J(A,B)$ is estimated by counting the number of indexes $i$-s, such that that
$h_{min}^{(i)}(A) = h_{min}^{(i)}(B)$, and this value is then divided by $k$. Observe that it holds that $h^{(i)}_{min}(A)=h^{(i)}_{min}(B)$ iff the minimum hash value of %
$A\cup B$ lies in $A\cap B$. 

This measure can be obtained by computing the cardinality of the intersection of $h_k(A)$ and $h_k(B)$ as follows. Each element $a_i$ of the vector $h_k(A)$ is encoded as $\langle a_i,i\rangle$. Similarly, $\langle b_i,i\rangle$ represents the $i$-th element of vector $h_k(B)$. An unbiased estimate of the Jaccard index of $A$ and $B$ is given by: %
\begin{equation}%
sim(A,B) = \dfrac{\left|\{\langle a_i,i\rangle\}_{i=1}^k \cap \{\langle b_i,i\rangle\}_{i=1}^k\right|}{k}%
\end{equation}

As discussed in~\cite{broder1998min}, if $\mathcal{F}$ is a family of min-wise independent hash functions,
then each value of a fixed set $A$ has the same probability to be the element with the smallest hash value, for all functions in $\mathcal{F}$. 
Specifically, for each min-wise independent hash function $h^{(i)}(\cdot)$ and for any set $S$,
we have that, for any $s_j,s_l\in S$, $\Pr[s_j = h^{(i)}_{min}(S)] = \Pr[s_l = h^{(i)}_{min}(S)]$. 
Thus, we also obtain that $Pr[h^{(i)}_{min}(A)=h^{(i)}_{min}(B)]=J(A,B)$.
In other words, if $r$ is a random variable that is 1 when $h^{(i)}_{min}(A) = h^{(i)}_{min}(B)$ and 0 otherwise, 
then $r$ is an unbiased estimator of $J(A,B)$; however, in order to reduce its variance, such random variable must be sampled several times, i.e., $k\gg1 $ hash values must be used.
In particular, by Chernoff bounds~\cite{chernoff1952measure}, the expected error of this estimate is $O(1/\sqrt{k})$~\cite{broder1997resemblance}.

\paragraph{Approximating (Jaccard) Similarity of Vectors without MinHash.}
If one needs to approximate the Jaccard index of two fixed-length {\em vectors} (rather than sets), 
one could use other (more efficient) techniques similar to MinHash. 
Observe that a vector $\overrightarrow S$ can be represented as a
set $S=\{\langle s_i,i\rangle\}$, where $s_i$ is simply the $i$-th element of  $\overrightarrow S$. 
We now discuss a new efficient strategy to approximate the Jaccard index of two vectors $A=\{\langle a_i,i\rangle\}_{i=1}^n$, $B=\{\langle b_i,i\rangle\}_{i=1}^n$ of length $n$, without using MinHash. 
Our approach incurs constant ($O(k)$) computational and communication complexity, i.e., it is independent from the length of the vectors being compared.

First, select $k$ random values $(r_1,\ldots,r_k)$, for $r_i$ uniformly distributed in $[1,n]$, and compute $A_k = \{\langle a_{r_i},r_i\rangle\}_{i=1}^k$ and $B_k = \{\langle b_{r_i},r_i\rangle\}_{i=1}^k$, respectively.
The value $\delta=|A_k\cap B_k|/k$ can then be used to assess the similarity of $A$ and $B$. %
We argue that $\delta$ is an unbiased estimate of $J(A,B)$: for each $\alpha \in (A_k \cup B_k)$ we have that $\Pr[\alpha \in (A_k \cap B_k)] = \Pr[\alpha \in (A \cap B)]$ since $\alpha \in (A \cap B) \wedge \alpha \in (A_k \cup B_k)\Leftrightarrow \alpha \in (A_k \cap B_k)$. We also have $\Pr[\alpha \in (A \cap B)] = J(A,B)$, thus, $\delta$
is indeed an unbiased estimate of $J(A,B)$.

The above algorithm implements a perfect min-wise permutation for this setting: since elements $(r_1, \ldots, r_k)$ are uniformly distributed, for each $i\in [1,k]$ any element in $A$ and $B$ has the same probability of being selected.
As such, similar to MinHash with many hash functions, the expected error is also $O(1/\sqrt{k})$.

\subsection{Cryptography Background}\label{subsec:psi-ca}

\paragraph{Private Set Intersection Cardinality (PSI-CA).} 
PSI-CA is a cryptographic protocol involving two parties: $\A$, on input $A \equal \{a_1,\ldots,a_w\}$,
and $\B$, on input $B \equal \{b_1,\ldots,b_v\}$, such that %
$\A$ outputs $|A\cap B|$, while $\B$ has no output. 
In the last few years, several PSI-CA protocols
have been proposed, including~\cite{ePrint,freedman2004efficient,kissner2005,vaidya2005}.

\paragraph{De Cristofaro et al.'s PSI-CA~\cite{ePrint}.}
Throughout this paper, we will use the PSI-CA construction presented by De Cristofaro, Gasti, and Tsudik in~\cite{ePrint}, denoted as \DGT\ in the rest of the paper,
as it offers the best communication and computation complexities (linear in set sizes).
\DGT\ is secure, in the presence of semi-honest adversaries,  under the Decisional Diffie-Hellman assumption (DDH) in the Random Oracle Model (ROM). 

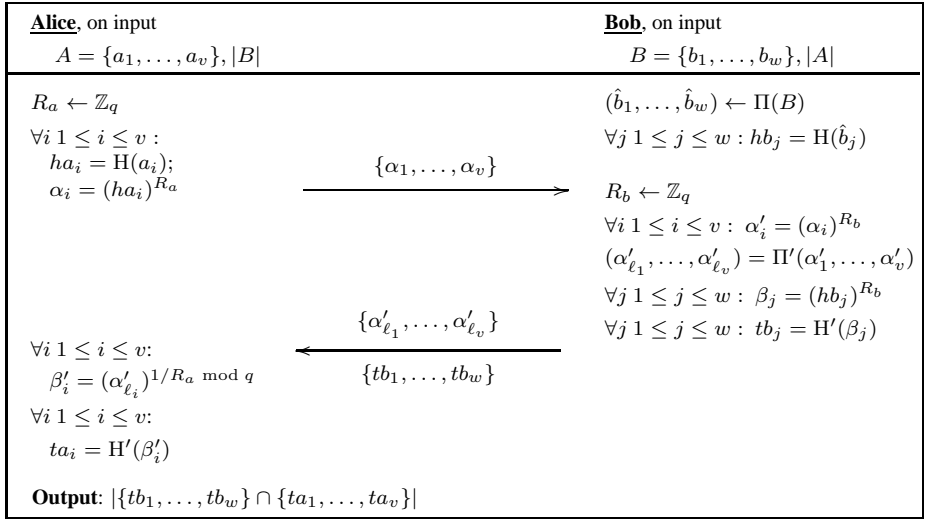
\begin{figure*}[t!]
\centering
{\fbox{\footnotesize
\begin{minipage}{0.72\columnwidth}
\begin{tabular}{lcl}
\textbf{\underline{Alice}}, on input 
&& \textbf{\underline{Bob}}, on input\vspace{0.1cm}\\
\hspace{0.25cm} $A=\{a_1,\ldots,a_v\},|B|$ & & \hspace{0.25cm} $B=\{b_1,\ldots, b_w\},|A|$ \vspace{-0.18cm}\\
\multicolumn{3}{l}{\hspace{-0.3cm}\line(1,0){344.5}} \vspace{0.1cm}\\
$R_a \leftarrow \Z_q $\vspace{0.1cm} && $(\hat b_1,\ldots, \hat b_w) \leftarrow \Pi (B)$\\ %
$\forall i\spa 1\leq i\leq v:$ && $\forall j\spa 1\leq j\leq w: hb_j = \mathrm{H}(\hat{b}_j)$\\
$\hspace{0.25cm} ha_i = \mathrm{H}(a_i); \; $ &&  \vspace{-0.45cm}\\

$\hspace{0.25cm} \alpha_i=({ha_i})^{R_a}$ &\hspace{0.0cm}  $\xymatrix@1@=100pt{\ar[r]^*+<4pt>{\{\alpha_1,\ldots,\alpha_v\}}&}$\vspace{-0.3cm}\\
& & $R_b \leftarrow \Z_{q}$\vspace{0.1cm}\\
&& $\forall i\spa 1\leq i\leq v:\spa \alpha_i' =(\alpha_i)^{R_b} $\vspace{0.1cm}\\
&& $(\alpha'_{\ell_1}, \ldots, \alpha'_{\ell_v}) = \Pi'(\alpha'_{1},\ldots,\alpha'_{v})$\vspace{0.1cm}\\
&& $\forall j\spa 1\leq j\leq w:\spa \beta_{j} ={(hb_{j})}^{R_b} $\vspace{0.1cm}\\
&& $\forall j\spa 1\leq j\leq w:\spa tb_{j} =\mathrm{H}'(\beta_{j})$\vspace{-0.6cm}\\
$\forall i\spa 1\leq i\leq v$: &\hspace{-0.2cm}  $\xymatrix@1@=100pt{& \ar[l]^*+<4pt>{\{tb_{1},\ldots,tb_{w} \}}_*+<4pt>{\{\alpha'_{\ell_1},\ldots,\alpha'_{\ell_v}\}}}$\vspace{-0.4cm}\\
\hspace{0.25cm}$\beta'_{i} = (\alpha'_{\ell_i})^{1/R_a \bmod q}$\vspace{0.1cm}\\
$\forall i\spa 1\leq i\leq v$:\vspace{0.1cm}\\
\hspace{0.25cm}$ta_{i} =\mathrm{H}'(\beta'_i)$\vspace{0.3cm}\\
\end{tabular}
{\color{white}P} {\bf Output}: $\left| \{tb_{1},\ldots,tb_{w}\}  \cap \{ta_1,\ldots,ta_v\} \right|$
\end{minipage}
}
}%
\vspace{-0.1cm}
\caption{\small PSI-CA protocol from~\cite{ePrint}, denoted as \DGT, executed on common input of two primes $p$ and 
$q$  (such that $q|p-1$), a generator $g$ of a subgroup of size $q$ and two hash functions, $\mathrm{H}$ and $\mathrm{H}'$, modeled as random oracles.  $\Pi(\cdot)$ and $\Pi'(\cdot)$ denote random permutations. {\em All computation is  mod $p$}.}
\label{fig:psi-ca}
\end{figure*}

The \DGT\ protocol is illustrated in Fig.~\ref{fig:psi-ca}. 
First, $\A$ hashes and masks its set items ($a_i$-s) with a random exponent ($R_a$) and sends resulting values ($\alpha_i$-s) to $\B$, which ``blindly'' exponentiates them with its own random value $R_b$. $\B$ then shuffles the resulting values ($\alpha'_i$-s) and sends them to $\A$.
Then, $\B$ sends $\A$ the output of a one-way function, $\mathrm{H}'(\cdot)$, computed over the exponentiations of $\B$'s (hashed and shuffled) items ($b_j$-s) to randomness $R_b$.
Finally, $\A$ tries to match one-way
function outputs received from $\B$ with one-way function outputs computed over
the shuffled ($\alpha'_i$-s) values, stripped of the initial randomness $R_a$. $\A$ learns the set intersection cardinality (and nothing else) by counting the number of such matches. Unless  they correspond to items in the intersection, one-way function outputs received from $\B$ cannot be used by $\A$ to learn related items in $\B$'s set (under the DDH assumption). Also, $\A$ does not learn {\em which} items are in the intersection as the matching occurs using  {\em shuffled} $\alpha_i$ values. 

\DGT\ requires $O(|A|+|B|)$ {\em offline}  and $O(|A|)$ {\em online} modular exponentiations in $\Z_p$ with exponents from subgroup $\Z_q$. (Offline operations are computed only once, for any number of interactions and any number of interacting parties).
Communication overhead amounts to $O(|A|)$ elements in $\Z_p$ and $O(|B|)$ -- in $\Z_q$. 
(Assuming 80-bit security parameter, $|q|=$ 160 bits and $|p|=$ 1024 bits.)

Protocol correctness is easily verifiable: for any $a_i$ held by \A\
and $b_j$ held by \B, if  $a_i=b_j$ (hence, $ha_i=hb_j$), we obtain:\vspace{-0.2cm}

$$
ta_{\ell_i}  =  \mathrm{H}'(\beta'_{\ell_i}) = \mathrm{H}'({\alpha_{\ell_i}}^{(1/R_a)})  =  
 \mathrm{H}'({ha_i}^{R_b}) =
 \mathrm{H}'({hb_j}^{R_b}) =
 \mathrm{H}'(\beta_j) =
 tb_{j} 
$$

\noindent Hence, \A\ learns set intersection cardinality by counting the number
of matching pairs $(ta_{\ell_i},tb_j)$.

\paragraph{Adversarial Model.} 
We use standard security models for secure two-party computation, which assume
adversaries to be either semi-honest or malicious.\footnote{Hereafter, 
the term {\em adversary} refers to protocol participants. Outside adversaries
are not considered, since their actions can be mitigated via standard network security techniques.}
As per definitions in~\cite{Goldreich}, protocols secure in the presence of
{\em semi-honest adversaries} assume that parties faithfully follow
all protocol specifications and do not misrepresent any information related to their inputs,
e.g., size and content. However, during or after protocol execution, any party might
(passively) attempt to infer additional information about other party's input.

Whereas, security in the presence of {\em malicious parties} allows  arbitrary  deviations
from the protocol. %
Security arguments in this paper are made with respect to \textbf{\em semi-honest} participants.

\section{Privacy-preserving Sample Set Similarity}\label{sec:new}
This section presents and analyzes our protocols for privacy-preserving computation
of sample set similarity, via secure computation 
of the Jaccard Similarity index. 

\subsection{Protocols Description}\label{sec:pj}
We propose two protocols -- both based on Private Set Intersection Cardinality (PSI-CA). The 
first protocol provides secure and exact computation of the Jaccard index, whereas, the other 
securely approximates it, using MinHash techniques, incurring significantly lower communication 
and computation overhead.

\begin{figure}[th]
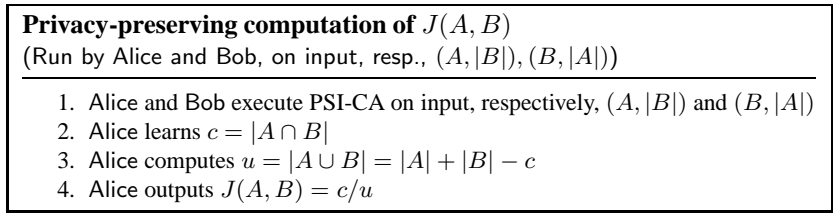

\begin{center}
\fbox{
\begin{minipage}{0.645\columnwidth}
{\textbf{Privacy-preserving computation  of ${J(A,B)}$}}\vspace{0.02cm}\\
{\small \sf (Run by $\A$ and $\B$, on input, resp., $(A,|B|),(B,|A|)$)}\vspace{-0.2cm}\\
\hspace{-0.35cm}\line(1,0){300} \vspace{-0.35cm}\\
{\small
\begin{compactenum}
\item $\A$ and $\B$ execute PSI-CA on input, respectively, $(A,|B|)$ and $(B,|A|)$
\item $\A$ learns $c = |A\cap B|$
\item $\A$ computes $u=|A\cup B| = |A|+|B|-c$
\item $\A$ outputs $J(A,B)=c/u$
\end{compactenum}}
\end{minipage}}
\vspace{-0.2cm}
\caption{Proposed protocol for privacy-preserving computation of set similarity.}
\label{fig:sjacc}
\vspace{-0.3cm}
\end{center}
\end{figure}

\paragraph{Privacy-Preserving Computation of Jaccard Index.} Fig.~\ref{fig:sjacc} illustrates 
our first protocol construction for securely computing the Jaccard index. It involves two 
parties, $\A$ and $\B$, on input sets $A$ and $B$, respectively, that wish to compute $J(A,B)$ 
in a privacy-preserving manner, i.e., in such a way that nothing is revealed about their input 
sets -- besides their size and joint Jaccard index. 

Given $|A|$, $|B|$ and $J(A,B)$, the size of the intersection between $A$ and $B$ can be easily 
computed as $|A\cap B| = J(A,B) \cdot (|A| + |B|) / (J(A,B) + 1)$. In other words, knowledge of $(|A|, |B|, J(A,B))$ is equivalent to knowledge of $(|A|, |B|, |A\cap B|)$. Therefore, in our protocol \A\ computes $|A\cap B|$ and uses it -- together with her input -- to derive $J(A,B)$.  As it is customary in secure-computation 
protocols, the size of parties' input is available to the counterpart, thus, it is included as 
part of the protocol input. The protocol does not assume any specific secure PSI-CA instantiation.

\paragraph{Privacy-Preserving Approximation of Jaccard Index using MinHash.} The computation of the Jaccard index, with or without privacy, can be relatively expensive when (1)
sample sets are very large, or (2) each set must be compared with a large number
of other sets -- since for each comparison, all computation must be re-executed {\em from scratch}.
Thus, MinHash is often used to 
estimate the Jaccard index, trading (bounded) error with appreciably faster computation.
In order to jointly and privately approximate $J(A,B)$ as 
$sim(A,B)$, \A\ and \B\ agree on $k$ and select a random subset of their sets using the MinHash 
technique in Section~\ref{subsec:minhash}. In particular, \A\ computes 
$\{\langle a_i,i\rangle\}_{i=1}^k$ where $a_i = h_{min}^{(i)}(A)$, and \B\ computes 
$\{\langle b_i,i\rangle\}_{i=1}^k$ for $b_i = h_{min}^{(i)}(B)$.
Similarity of two sample sets is then computed as $sim(A,B) = |\{\langle a_i,i\rangle\}_{i=1}^k 
\cap \{\langle b_i,i\rangle\}_{i=1}^k|/k$ using any secure instantiation of PSI-CA.
Therefore, privacy-preserving estimation of the Jaccard index, using multi-hash MinHash, can be
reduced to securely computing cardinality of set intersection. 
The resulting protocol is presented in Fig.~\ref{fig:spmin} below.

\begin{figure}[ht]
\begin{center}
\fbox{
\begin{minipage}{0.73\columnwidth}
{\textbf{Private Jaccard index estimation ${sim(A,B)}$}}\vspace{0.02cm}\\
{\small{ Run by $\A$ and $\B$, on common input $k$ and private input $\{\langle a_i,i\rangle\}_{i=1}^k$ (for \A) and $\{\langle b_i,i\rangle\}_{i=1}^k$} (for \B)}\vspace{-0.2cm}\\
\hspace{-0.35cm}\line(1,0){343} \vspace{-0.35cm}\\
{\small
\begin{compactenum}
\item $\A$ and $\B$ execute PSI-CA on input, resp., $(\{\langle a_i,i\rangle\}_{i=1}^k, k)$ and
$(\{\langle b_i,i\rangle\}_{i=1}^k,k)$
\item $\A$ learns $\delta=|\{\langle a_i,i\rangle\}_{i=1}^k\cap\{\langle b_i,i\rangle\}_{i=1}^k|$
\item $\A$ outputs $sim(A,B)= \delta/k$
\end{compactenum}}
\end{minipage}}
\vspace{-0.2cm}
\caption{Proposed protocol for privacy-preserving approximation of set similarity.\vspace{0.1cm}}
\label{fig:spmin}
\vspace{-0.2cm}
\end{center}
\end{figure}

It is easy to observe that, compared to the Jaccard index computation (Fig.~\ref{fig:sjacc}), the use of MinHash leads 
to executing PSI-CA on smaller sets, as $k \ll {\rm Min}(|A|, |B|)$. %
Communication and computation overhead only depend on $k$, since inputs to PSI-CA are now sets of $k$ items.

\subsection{Security Analysis}

\paragraph{Security of Privacy-Preserving Computation of Jaccard Index.} 
Informally, by running the protocol in Fig.~\ref{fig:sjacc}, parties do not reciprocally 
disclose the content of their private sets. \A\ learns similarity computed as the 
Jaccard index, while both parties learn the size of the other party's input.

The security of the protocol in Fig.~\ref{fig:sjacc} relies on the security of the 
instantiation of the underlying PSI-CA protocol. In particular we assume that, in the semi-
honest model, PSI-CA only reveals $|A\cap B|$ to \A, while \B\ learns nothing besides $|A|$.

$\A$ and $\B$ do not exchange any information besides messages related to the PSI-CA protocol. 
For this reason, a secure implementation of the underlying PSI-CA guarantees that neither $\A$ 
nor $\B$ learn additional information about the other party's set.
Since knowledge of $(|A|, |B|, J(A,B))$ is equivalent to knowledge of $(|A|, |B|, |A\cap B|)$, 
the protocol in Fig.~\ref{fig:sjacc} is secure in the semi-honest setting.

\paragraph{Security of Privacy-Preserving Approximation of Jaccard Index.} Similar to the 
protocol in Fig.~\ref{fig:sjacc}, the security of protocol in Fig.~\ref{fig:spmin} 
relies on the security of the underlying PSI-CA construction. In particular, $sim(A,B)$ is 
defined as the size of the intersection between $\{\langle a_i,i\rangle\}_{i=1}^k$ and 
$\{\langle b_i,i\rangle\}_{i=1}^k$, divided by a (public) constant $k$. Therefore, any 
information \A\ and \B\ learn about the other party's input can also be used to break the 
underlying PSI-CA protocol. Since the PSI-CA protocol is assumed to be secure, \A\ and \B\ do 
not learn additional information.

$k$ is selected independently from $|A|$ and $|B|$, therefore it does not reveal any 
information about the two sets. PSI-CA, together with the way input is constructed, conceals 
the relationship between $k$ and $|A|,|B|$ by 
not disclosing how many elements $a_i=a_j$ and $b_i=b_j$ for $i\neq j$ on the parties' inputs. 
Therefore, the protocol does not disclose the size of \A\ and \B's inputs.

\paragraph{Extension of Privacy-Preserving Approximation of Jaccard Index.}
In the previous protocol, $\A$ learns some additional information compared to the protocol in 
Fig.~\ref{fig:sjacc}. In particular, rather than computing the similarity -- and therefore the 
size of the intersection -- of sets $A$ and $B$, she determines how many elements from a 
particular subset of $A$ (constructed using MinHash) also appear in the subset selected from 
$B$. We now provide a brief overview of how this issue can be fixed efficiently.

$\A$ and $\B$ can construct their input sets (i.e., $\{\langle a_i,i\rangle\}_{i=1}^k$ and $\{\langle b_i,i\rangle\}_{i=1}^k$) using a set of Oblivious Pseudo Random Function (OPRF) evaluations\footnote{An Oblivious Pseudo Random Function (OPRF) is a two-party protocol, involving one party, on input $key$, and another, on input $s$. At the end of the interaction the former learns nothing, while the latter obtains $f_{key}(s)$, where $f$ is a pseudo-random function.} rather than a set of hash functions: $\A$ and $\B$ engage in a multi-party protocol where $\A$ inputs her set $A = \{a_1,\ldots,a_v\}$ and learns a random permutation of OPRF$_{key_j}(a_1),\ldots,$OPRF$_{key_j}(a_v)$ for $1\leq j \leq k$. $\A$ constructs her input selecting the smallest value OPRF$_{key_j}(a_i)$ for each $j$. $\B$ constructs his input without interacting with $\A$. 
While the cost of this protocol is linear in the size of the input sets, it is significantly higher than that of protocol Fig.~\ref{fig:spmin}.

\subsection{Performance Evaluation}

\subsubsection{Privacy-Preserving Computation of Jaccard Index.}
Cost of protocol in Fig.~\ref{fig:sjacc} is dominated by that incurred by the underlying PSI-CA protocol. While it could be instantiated using {\em any} PSI-CA construction, we choose \DGT\ in order to maximize efficiency. This protocol, reviewed in Section~\ref{subsec:psi-ca}, incurs linear communication and computational complexities, thus, overall complexities of protocol in Fig.~\ref{fig:sjacc} are also linear in the size of sets. If we were to compute the Jaccard index without privacy,
asymptotic complexities would be same as our privacy-preserving protocol -- i.e., linear. 
However, given the lack of cryptographic operations,
constants hidden by the big $O(\cdot)$ notation  would be much smaller. %

To assess the real-world practicality of resulting construction, protocol in Fig.~\ref{fig:sjacc} has  been implemented in C (with OpenSSL and GMP libraries), using 160-bit random exponents and 1024-bit moduli to obtain 80-bit security.
We run experiments on sets such that ${|A|=|B|=1000}$. Recall that, in \DGT\ items are always hashed (\DGT\ assumes ROM), so their size is non-influent. We use select SHA-1 as the hash function, thus, hashed items are 160-bit. 

In this setting, protocol in Fig.~\ref{fig:sjacc} incurs (i) $\mathbf{0.5s}$ total computation time
on a single Intel Xeon E5420 core running at 2.50GHz
and (ii) $\mathbf{276}$KB in bandwidth. We omit running times for larger sets since, as complexities are linear,
one can easily derive a meaningful estimate of time/bandwidth for virtually any size.

We also implement an optimized prototype that further improves total running time
by (1) pipelining computation and transmission and (2) parallelizing computation on two cores. 
We test the prototype by running 
$\A$ and $\B$ on two PCs equipped with 2 quad-core Intel Xeon E5420 processors
running at 2.50GHz, however, we always use (at most) 2 cores.
On a conservative stance, we do not allow parties to perform any pre-computation offline.
We simulate a 9Mbps link, since, according to~\cite{49},
it represents the current average Internet bandwidth in US and Canada.
In this setting, and again considering $|A|=|B|=1000$,
total running time of protocol in Fig.~\ref{fig:sjacc} amounts to $\mathbf{0.23s}$.
Whereas, the computation of Jaccard index {\em without privacy}
takes $\mathbf{0.018s}$. 
Therefore, we conclude that privacy protection, in our experiments,
only introduces a (roughly) \textbf{12-fold slowdown}, independently from set sizes.

\paragraph{Comparison to prior work.}
Performance evaluation above does not include any prior solutions, since,
to the best of our knowledge, there is no comparable cryptographic primitive
for privacy-preserving computation of the Jaccard index. 
The work in~\cite{singh2009privacy} is somewhat related:
it targets private computation of the Jaccard index using Private Equality Testing (PET)~\cite{lipmaa2003verifiable}
and deterministic encryption. However, it introduces the need for a non-colluding semi-honest {\em third party}, which violates our design model. Also, it incurs an impractical number of public-key operations, 
i.e., {\em quadratic} in the size of sample sets (as opposed to linear in our case).
Finally, additional (only vaguely) related techniques include: (i) work on privately approximating dot product of two vectors, such as,~\cite{rulemining,PSDM04}, and (ii) probabilistic/approximated private set operations based on Bloom filters, such as,~\cite{rulemining,kerschbaum2012outsourced}. (None of these techniques, however, can be used to
solve problems considered in this paper.)

\subsubsection{Privacy-Preserving Estimation of Jaccard Index with MinHash}
We also tested the performance of our construction for privacy-preserving {\em approximation} of Jaccard similarity, again, using \DGT, i.e., the PSI-CA from~\cite{ePrint}.
Once again, we select sets with 1000 items, 1024-bit moduli and 160-bit random exponents, and run experiments on two PCs with 2.5GHz CPU and a 9Mbps link. We select $\mathbf{k=400}$ in order to have an estimated error of about $5\%$. (Recall that, as mentioned in Section~\ref{subsec:minhash} the error is approximated as $1/\sqrt{k}$). 

In this setting, the total running time of protocol in Fig.~\ref{fig:spmin} amounts to $\mathbf{0.09s}$ -- less than half compared to the one in Fig.~\ref{fig:sjacc}.
Whereas, in the same setting, the approximation of Jaccard index {\em without privacy} takes $\mathbf{0.007s}$. Thus,
the slow-down factor introduced by the privacy-protecting layer (similar to the protocol proposed in Section~\ref{sec:pj}) is {\bf 12-fold}.
Again, note that times for different values of $k$ can be easily estimated since the complexity of the protocol is linear.

\paragraph{Comparison to Prior Work.}
The estimation of set similarity through MinHash -- whether privacy-preserving
or not -- requires counting the number of times for which it holds that
$h_{min}^{(i)}(A)=h_{min}^{(i)}(B)$, with $i=1,\ldots,k$. We have denoted this number as $\delta$.
Protocol in Fig.~\ref{fig:spmin} above attains secure computation of $\delta$ through privacy-preserving
set intersection cardinality.
However, it appears somewhat more intuitive to do so by using the approach proposed
by~\cite{pino} in the context of social-network friends discovery.
Specifically, in~\cite{pino}, $\A$ and $\B$ compute, resp., $\{a_i\}_{i=1}^k$ and $\{b_i\}_{i=1}^k$, just like in our protocol.
Then, $\A$ generates a public-private keypair $(pk,sk)$ for Paillier's additively homomorphic encryption
cryptosystem~\cite{Pa99} and sends $\B$ $\{z_i = Enc_{pk}(a_i)\}_{i=1}^k$.
$\B$ computes $\{(z_i\cdot Enc_{pk}(-b_i))^{r_i}\}_{i=1}^k$ for random $r_i$'s and
returns the resulting vector of ciphertexts after shuffling it. Upon decryption, $\A$ learns
$\delta$ by counting the number of $0$'s. %
Nonetheless, the technique proposed by~\cite{pino} actually incurs
an increased complexity, compared to our protocol in Fig.~\ref{fig:spmin} 
(instantiated with \DGT).
Assuming 80-bit security parameters, thus, 1024-bit moduli and
160-bit subgroups, and 2048-bit Paillier moduli,
and using \mmm\ to denote a multiplication of 1024-bit numbers,
multiplications of 2048-bit numbers
count for 4\mmm. Using square-and-multiply, exponentiations with $q$-bit exponents modulo 1024-bit count for $(1.5|q|)\mmm$.
In~\cite{pino}, $\A$ performs $k$ Paillier encryptions (i.e., $2k$ exponentiations and $k$ multiplications) and $k$ decryptions (i.e., $k$ exponentiations and multiplications), while $\B$ computes $k$ exponentiations and multiplications. Therefore, the total computation complexity amounts to $(6\cdot4\cdot1.5\cdot1024+4\cdot4)\mmm k = 36,880\mmm$.
Whereas, our approach (even without pre-computation)
requires both $\A$ and $\B$ to perform $4k$ exponentiations of 160-bit numbers modulo 1024-moduli
and $2k$ multiplications, i.e., $(4\cdot1.5\cdot160+2)k\mmm = 962 \mmm k$, thus, our protocol
achieves a 38-fold efficiency improvement.
Communication overhead is also higher in~\cite{pino}: it amounts to $(2\cdot2048)k$ bits; whereas,
using PSI-CA, we need to transfer $(2\cdot1024+160)k$ bits, i.e., slightly more than half the traffic.

\section{Privacy-Preserving Document Similarity}\label{sec:docsimilarity}

After building efficient (linear-complexity) primitives for privacy-preserving
computation/approximation of Jaccard index, we now explore their applications
to a few compelling problems. %
We start with evaluating the similarity of two documents, which is relevant in
many common applications, including copyright protection, file management, plagiarism prevention,
duplicate submission detection, law enforcement.
In last few years, the security community has started investigating privacy-preserving techniques to enable
detection of similar documents without disclosing documents' actual content. Below, we first review prior work and, 
then, present our technique for efficient privacy-preserving document similarity.

\subsection{Prior Work}
The work in~\cite{jiang2008similar} (later extended in~\cite{MurugesanVLDB2010})
is the first to realize {\em privacy-preserving document similarity}. 
It realizes secure computation of the \emph{cosine similarity} 
of vectors representing the documents, i.e., each document is represented
as the list of words appearing in it, along with the normalized number of occurrences.
Recently, Jiang and Samanthula~\cite{JiSa} have proposed a novel technique
relying on the Jaccard index and $N$-gram based document representation~\cite{manber1994finding}. 
(Given any string, an $N$-gram is a substring of size $N$).
According to~\cite{JiSa}, the $N$-gram based technique presents several advantages over 
cosine similarity: (1) it improves on finding {\em local similarity}, e.g., overlapping of 
pieces of texts, (2) it is language-independent, (3) it requires a much simpler 
representation, and (4) it is less sensitive to document modification. We overview it below.

\begin{figure}[ht]
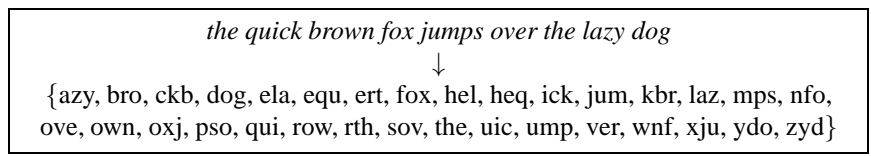

\centering
\fbox{
\begin{tabular}{c}
{\em the quick brown fox jumps over the lazy dog}\\
$\downarrow$\\
\{azy, bro, ckb, dog, ela, equ, ert, fox, hel, heq, ick,
jum, kbr, laz, mps, nfo,\\
 ove, own, oxj, pso, qui, row,
rth, sov, the, uic, ump, ver, wnf, xju, ydo, zyd\}
\end{tabular}
}
\vspace{-0.2cm}
\caption{Tri-gram representation.}\label{fig:fox}
\end{figure}

\paragraph{Documents as sets of {\em N}-grams.} A document can be represented as a set of $N$-grams contained in it.
To obtain such a representation, one needs to remove spaces and punctuation
and build the set of successive $N$-grams in the document.
An example of a sentence, along with its $N$-gram representation (for $N=3$), is illustrated in Fig.~\ref{fig:fox}.
The similarity of two documents can then be estimated as the {\em Jaccard index} of the two 
corresponding sets of $N$-grams. 
In the context of document similarity, experts point out that 3 results 
as a good choice of $N$~\cite{broder1997resemblance}.

To enable privacy-preserving computation of Jaccard index, and therefore %
estimation of document similarity, Jiang and Samanthula~\cite{JiSa} propose a two-stage protocol based
on {\em Paillier}'s additively homomorphic encryption~\cite{Pa99}.
Suppose $\A$ wants to privately evaluate the similarity of her document $D_A$ against
a list of $n$ documents held by $\B$, i.e., $D_{B:1}, \ldots, D_{B:n}$.
First, $\B$ generates a global space, $|S|$, of tri-grams based on his document collection.
This way, $D_A$ as well as each of $\B$'s document, $D_{B:i}$, for $i=1,\ldots,n$, can be represented as binary vectors
in the global space of tri-grams: each component is 1 if the corresponding tri-gram
is included in the document and 0 otherwise. In the following, we denote with $A$ the representation of $D_A$ and
with $B_i$ that of $D_{B:i}$.

Then, $\A$ and $\B$ respectively compute random shares
$a$ and $b_i$ such that $a+b_i = |A\cap B_i|$. Next, they set $c = |A| - a$ and $d_i = |B|-b_i$.
Finally, $\A$ and $\B$, on input $(a,c)$ and
$(b_i,d_i)$, resp., execute a Secure Division protocol (e.g.,~\cite{bunn2007secure,AtallahBLFT04}) 
to obtain $(a+b_i)/(c+d_i) = |A\cap B_i|/|A\cup B_i| = J(A,B_i)$.

The computational complexity of the protocol in~\cite{JiSa}
amounts to $O(|S|)$ Paillier encryptions performed by $\A$, and $O(n\cdot |S|)$ modular multiplications
-- by $\B$. Whereas, communication overhead amounts to $O(n\cdot |S|)$ Paillier ciphertexts. %

\paragraph{Flaw in~\cite{JiSa}.} Unfortunately, protocol in \cite{JiSa} {\em is not secure}, since
$\B$ has to disclose his global space of tri-grams (i.e., the set of all tri-grams appearing in his document 
collection). Therefore, $\A$ can passively check whether or not a word appears in $\B$'s document 
collection. Actually, $\A$ can learn much more, as we show in Appendix~\ref{app:flaw}.
We argue that this flaw could be fixed by considering the global space of tri-grams
as the set of all possible tri-grams, thus, avoiding the disclosure of $\B$'s tri-grams set.
Assuming that documents are stripped of any symbol and contain
only lower-cased letters and digits, we obtain $S=\{a,b,\ldots,z, 0,1,\ldots,9\}^3$.
Unfortunately, this modification would tremendously increase computation and communication
overhead. %

\subsection{Our Construction}
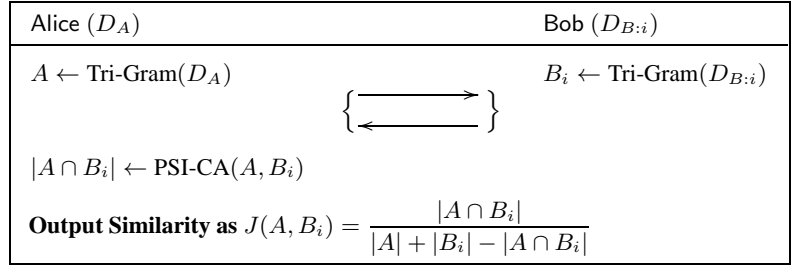
\begin{figure}[bth]
\centering
\hspace{1cm}
\fbox{\small
\begin{minipage}{0.61\columnwidth}
\begin{tabular}{lcl}
\hspace{-0.05cm}$\A$ $(D_A)$ & &\hspace{0.3cm} $\B$ $(D_{B:i})$ \vspace{-0.2cm}\\
\multicolumn{3}{c}{\hspace{-0.3cm}\line(1,0){293.5} \vspace{0.1cm}}\\
\hspace{-0.05cm}$A\leftarrow\mbox{Tri-Gram}(D_A)$ & & \hspace{0.3cm} $B_i\leftarrow\mbox{Tri-Gram}(D_{B:i})$\hspace{0.2cm}\\
\hspace{3.99cm} \Big\{ \Big. \vspace{-0.69cm} & & \hspace{-0.55cm} \Big. \Big\}\\
&\hspace{-0.62cm}$\xymatrix@1@=45pt{\ar[r]^*+<4pt>{}&}$\vspace{0.02cm}\\
 &\hspace{-0.62cm}$\xymatrix@1@=45pt{& \ar[l]^* +<4pt>{}_*+<4pt>{}}$ &  \vspace{-0.2cm}\\
\\
\hspace{-0.05cm}$|A\cap B_i| \leftarrow \mbox{PSI-CA}(A,B_i)$\vspace{-0.05cm}\\
\end{tabular}
\vspace{0.2cm}

\hspace{0.14cm}{\bf Output Similarity as} %
$J(A,B_i)=\dfrac{|A\cap B_i|}{|A|+|B_i|-|A\cap B_i|}$ %
\end{minipage}
}
\caption{Privacy-preserving {\em evaluation} of document similarity of documents $D_A$ and $D_{B:i}$.}
\label{fig:ppds}
\end{figure}

As discussed in Section~\ref{sec:new}, we can realize privacy-preserving
computation of the Jaccard index using PSI-CA. 
To privately evaluate the similarity of documents $D_A$ and any document $D_{B:i}$, $\A$ and
$\B$ execute protocol in Fig.~\ref{fig:ppds}. Function $\mbox{Tri-Gram}(\cdot)$
denotes the representation of a document as the set of tri-grams appearing in it.

\paragraph{Complexity.} Complexity of protocol in Fig.~\ref{fig:ppds}
is bounded by that of the underlying PSI-CA construction.
Using \DGT, computational complexity amounts to $O(|A|+|B_i|)$ modular
exponentiations, whereas, communication overhead -- to $O(|A|+|B_i|)$. %
Observe that, in the setting where $\A$ holds one documents and $\B$ a collection of $n$ documents,
complexities should be amended to $O(n|A|+\sum_{i=1}^n |B_i|)$. 
However, due to the nature of \DGT, $\B$ can perform $O(\sum_{i=1}^n |B_i|)$ computation {\em off-line}, ahead of time.
Hence, total {\em online} computation amounts to $O(n|A|)$.

\begin{figure}[t!]
\centering
\hspace{1cm}
\fbox{\small
\begin{minipage}{0.62\columnwidth}
\begin{tabular}{lcl}
\hspace{-0.05cm}$\A$ $(D_A)$ & &\hspace{-3.8cm} $\B$ $(D_{B:i})$ \vspace{-0.15cm}\\
\multicolumn{3}{c}{\hspace{-0.5cm}\line(1,0){299} \vspace{0.1cm}}\\
\hspace{-0.05cm}$A\leftarrow\mbox{Tri-Gram}(D_A)$ & & \hspace{-3.8cm} $B_i\leftarrow\mbox{Tri-Gram}(D_{B:i})$\vspace{0.1cm}\\
\hspace{-0.05cm}$h_k(A)\leftarrow\mbox{MinHash}(A)$ & & \hspace{-3.8cm} $h_k(B_i)\leftarrow\mbox{MinHash}(B_i)$\hspace{-0.35cm}\spa\vspace{0.1cm}\\
\hspace{3.95cm} \Big\{ \Big. \vspace{-0.7cm} & & \hspace{-4.55cm} \Big. \Big\}\\
&\hspace{-9.6cm}$\xymatrix@1@=45pt{\ar[r]^*+<4pt>{}&}$\vspace{-0.0cm}\\
 &\hspace{-9.6cm}$\xymatrix@1@=45pt{& \ar[l]^*+<4pt>{}_*+<4pt>{}}$ &  \vspace{-0.25cm}\\
\hspace{-0.05cm}$|h_k(A)\cap h_k(B_i)|\leftarrow$\vspace{0.05cm}\\
$\;\;\;\;\;\mbox{PSI-CA}(h_k(A),h_k(B_i))$\vspace{0.15cm}\\
\hspace{-0.05cm}{\bf Output Similarity Approximation as}: %
$sim(A,B_i)=\dfrac{|h_k(A)\cap h_k(B_i)|}{k}$ %
\end{tabular}
\end{minipage}
}
\vspace{-0.2cm}
\caption{Privacy-preserving {\em approximation} of document similarity of documents $D_A$ and $D_{B:i}$.}
\label{fig:app-ppds}
\vspace{-0.2cm}
\end{figure}
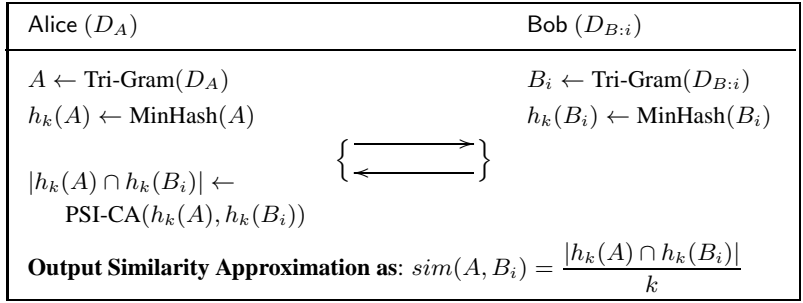

\paragraph{More efficient computation using MinHash.} As discussed in Section~\ref{subsec:minhash}, one can approximate
the Jaccard index by using MinHash techniques, thus, trading off accuracy with significant improvement in protocol
complexity. The resulting construction is similar to the one presented above and is illustrated
in Fig.~\ref{fig:app-ppds}. It adds an intermediate step between the tri-gram representation and the 
execution of PSI-CA: $\A$ and $\B$ apply MinHash to sets $A$ and $B_i$, respectively, and obtain $h_k(A)$ and $h_k(B_i)$.
The main advantage results from the fact that PSI-CA is now executed on smaller sets, of constant size $k$, 
thus, achieving significantly improved communication and computational complexities.
Again, note that the error is bounded by $O(1/\sqrt{k})$.

\begin{table}[b!]
{\small 
\centerline{
\begin{tabular}{|c|c|c|c|c|} \hline
\multirow{2}[2]{*} {$n$} & 
\multirow{2}[2]{*} {\cite{JiSa}} & 
\multirow{2}[2]{*} {Fig.~\ref{fig:ppds}} &  
\multicolumn{2}{c|}{Fig.~\ref{fig:app-ppds}} \\ \cline{4-5}
& & & $k=100$ & $k=40$\\ \hline
$10$   & 9.5 mins  & 6.3 secs  & 0.05 secs & 0.05 secs\\
$10^2$ & 9.9 mins  & 63 secs   & 1.9 secs & 1.9 secs\\
$10^3$ & 12.7 mins & 10.4 mins    & 48 secs & 19.2 secs\\
$10^4$ & 40.7 mins & 1.74 hours & 8 mins & 3.2 mins\\
$10^5$ & 5.3 hours & 17.4 hours & 1.2 hours & 32 mins \\
\hline
\end{tabular}
\vspace{-0.2cm}
}
\caption{Computation time of privacy-preserving document similarity. $n$ denotes the number of documents.}
\label{tab:expp}
}
\end{table}

\paragraph{Performance Evaluation.}
We now compare the performance of our constructions to the most efficient prior technique, i.e., the protocol
in~\cite{JiSa} (that, unfortunately, is insecure).
We consider the setting of~\cite{JiSa}, where $\B$ maintains a collection of $n$ documents.
Recall that our constructions use \DGT, i.e., the PSI-CA in~\cite{ePrint}. Assuming 80-bit security parameters,
we select 1024-bit moduli and 160-bit random exponents. As \cite{JiSa} relies on Paillier encryption,
it uses 2048-bit moduli and 1024-bit exponents. %
In the following, let \mmm\ denote a multiplication of 1024-bit numbers. Multiplications of 2048-bit numbers
count for 4\mmm. Modular exponentiations with $q$-bit exponents modulo 1024-bit count for $(1.5|q|)\mmm$.
The protocol in~\cite{JiSa} requires $O(|S|)$ Paillier encryptions and $O(n\cdot|S|)$ modular multiplications.
We need $|S|=36^3=46,656$ as we consider 3-grams and 26 alphabet letters plus [0--9] digits.
Therefore, the total complexity amounts to $(4\cdot 1.5\cdot 1024 + 4n)|S|\mmm = (6144+4n)|S|\mmm \approx
(2.9\cdot 10^8 + 1.9\cdot 10^5n)\mmm$.

Our construction above requires $(2\cdot1.5\cdot160n(max(|A|,\{B_i\}_{i=1}^n))\mmm$
for the computation of Jaccard index similarity and ($1.5\cdot160nk)\mmm$ for its approximation.
Thus, to compare performance of our protocol to that of~\cite{JiSa}, we need to take into account the dimensions of $A$, $B_i$, as well as $n$ and $k$. 
To this end, we collected 393 scientific papers from the KDDcup dataset of scientific papers published
in ArXiv between 1996 and 2003~\cite{kdd}. The average number of different tri-grams
appearing in each paper is 1307.
Therefore, cost of our two techniques can be estimated as $(2\cdot1.5\cdot160\cdot1307n)\mmm$
and $(1.5\cdot160\cdot nk)\mmm$, respectively.
Thus, our technique for privacy-preserving document similarity is faster than~\cite{JiSa}
for $n<2000$. Furthermore, using MinHash techniques,
complexity is always faster (and of at least one order of magnitude), using both $k=40$ and $k=100$. 
Also, recall that, as opposed to ours, the protocol in~\cite{JiSa} is {\bf\em not} secure.

Assuming that it takes about 1$\mu s$ to perform modular multiplications of 1024-bit integers (as per our experiments
on a single Intel Xeon E5420 core running at 2.50GHz),
we report estimated running times in Table~\ref{tab:expp}
for increasing values of $n$ (i.e., the number of $\B$'s documents).

We performed some statistical analysis to determine the real magnitude of the error 
introduced by MinHash, when compared to the Jaccard index without MinHash.
Our analysis is based on the trigrams from documents in the KDDcup dataset \cite{kdd}, and confirms that the average error is within the expected bounds: for $k=40$, we obtained an average error of 14\%, while for $k=100$ the average error was 9\%.
This is acceptable, considering that the Jaccard index actually provides a normalized {\em estimate} of the similarity 
between two sets, not a definite metric.

\section{Privacy-Preserving Iris Matching}\label{sec:iris}
Advances in biometric recognition enable the use of biometric data 
as a practical mean of authentication and identification. Today,
several governmental agencies around the world perform large-scale collections of different biometric features.
As an example, the US Department of Homeland Security (DHS) collects face, fingerprint and iris images,
from visitors within its US-VISIT program~\cite{usvisit}. Iris images are also collected 
from all foreigners, by the United Arab Emirates (UAE) Ministry of Interior, as well as fingerprints and photographs 
from certain types of travelers~\cite{uaei}.

While biometry serves as an excellent mechanism for
identification of individuals, biometric data is, undeniably,
extremely sensitive and must be subject to minimal exposure. 
As a result, such data cannot be disclosed arbitrarily.  %
Nonetheless, there are many legitimate scenarios where biometric data
should be shared, in a controlled way, between different entities. 
For instance, an agency may need to determine whether a given biometric appears on a government watch-list.
As agencies may have different clearance levels, privacy of biometric's owner should be preserved if no matches are found, but, at the same time, unrestricted access to the watch-list cannot be granted.

\subsection{Prior Work}
As biometric identification techniques are increasingly employed,
related privacy concerns have been investigated by the research community~\cite{CimatoGPSS09}.
A number of recent results address the problem of
privacy-preserving face recognition. The work in~\cite{erk09} is the first to present
a secure protocol, based on Eigenfaces, later improved by~\cite{sad09}.
Next,~\cite{scifi} designs a new privacy-preserving  face recognition algorithm, called SCiFI. %
Furthermore, the protocol in~\cite{bar10} realizes privacy-preserving fingerprint identification, using FingerCodes~\cite{jai00}. FingerCodes use texture information from a fingerprint to compare two biometrics. The algorithm is not as discriminative as traditional fingerprint matching techniques based on location of minutiae points, but it is adopted
in~\cite{bar10} given its suitability for efficient {\em privacy-preserving} realization. %
Among all biometric techniques, this paper focuses on {\em iris-based} identification.
The problem of privacy-preserving iris matching has been  introduced by Blanton and Gasti
in~\cite{Blanton11esorics}, who propose an approach
based on a combination of garbled circuits~\cite{yao86} and homomorphic encryption.

\subsection{Our Construction}
A (human) iris can be digitalized as an $n$-bit string $S=s_1s_2\cdots s_n$ with an $n$-bit mask 
$M_S = ms_1ms_2\cdots ms_n$. The mask indicates which bits of $S$ have been read reliably. 
In particular, the $i$-th bit of $S$ should be used for matching only if the $i$-th bit of $M_S$ is set to 1.
A common value for $n$, which we use in our experiments, is 2048.
As, during an iris scan, the subject may rotate its head, a right or left shift can occur
in the iris representation, depending on the direction of the rotation. 
Therefore, the distance between two irises $A$ and $B$ is computed as the minimum distance between all rotations of $A$ and the representation of $B$.
In practice, it is reasonable to assume that the rotation is limited to a shift of at most 5 positions towards left/right \cite{Blanton11esorics}.

The matching between two irises, $A$ and $B$, is computed via the {\em Weighted Hamming Distance} ($\mathrm{WHD}$) 
of the samples.
Let $M = (M_A \wedge M_B)$; $\mathrm{WHD}$ is computed as:\vspace*{-0.1cm}
\begin{equation} \label{eq:hd}
\mathrm{WHD}(A, B, M) = \frac{HD(A \wedge M, B \wedge M) } {\|M\|}
\vspace{-0.1cm}
\end{equation}
where $\| \cdot \|$ denotes hamming weight, i.e. the number of string bits set to 1.
Given a threshold $t$, if $\mathrm{WHD}(A, B, M)<t$, we say that irises $A$ and $B$ are 
{\em matching}. (Assuming a maximum rotation of 5 positions, the distance must be computed 11 times.)

In the following, we propose a probabilistic technique for privately
estimating of $\mathrm{WHD}(A,B,M)$, that relies on the construction for privacy-preserving estimation of Jaccard index based on MinHash (introduced in Section~\ref{sec:pj}).
The error on the approximation is bounded by the MinHash parameter $k$.

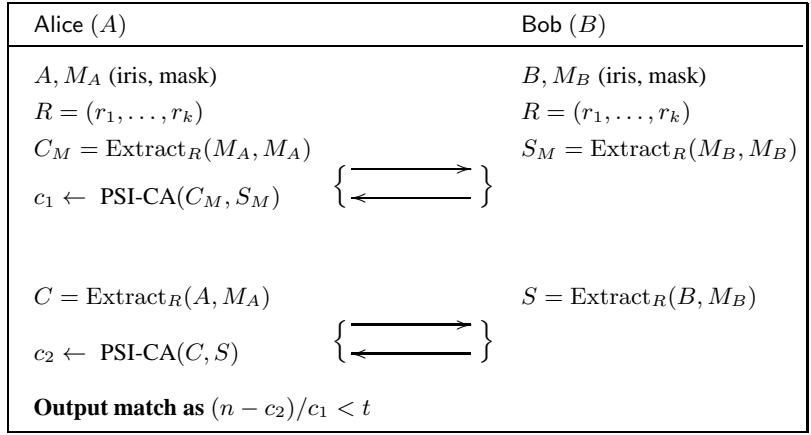
\begin{figure}[ttt!]
\centering
\fbox{\small
\centering
\begin{minipage}{0.625\columnwidth}
\begin{tabular}{lcl}
\hspace{0.02cm}$\A$ $(A)$ & &\hspace{0.1cm} $\B$ $(B)$ \vspace{-0.2cm}\\
\multicolumn{3}{c}{\hspace{-0.3cm}\line(1,0){300} \vspace{0.1cm}}\\
\hspace{0.02cm}$A, M_A$ (iris, mask) & & \hspace{0.1cm} $B,M_B$ (iris, mask)\vspace{0.1cm}\\
\hspace{0.02cm}$R={(r_1, \ldots, r_k)}$ & & \hspace{0.1cm} $R={(r_1, \ldots, r_k)}$ \vspace{0.1cm}\\

\hspace{0.02cm}$C_M = \mathrm{Extract}_R(M_A, M_A)$ & & \hspace{0.1cm} $S_M=\mathrm{Extract}_R(M_B, M_B)$\hspace{-0.35cm}\\
\hspace{3.9cm} \Big\{ \Big. \vspace{-0.7cm} & & \hspace{-0.55cm} \Big. \Big\}\\
&\hspace{-0.6cm}$\xymatrix@1@=45pt{\ar[r]^*+<4pt>{}&}$\vspace{-0.0cm}\\
\hspace{0.02cm}$c_1 \leftarrow \;\mbox{PSI-CA}(C_M,S_M)$ &\hspace{-0.6cm}$\xymatrix@1@=45pt{& \ar[l]^*+<4pt>{}_*+<4pt>{}}$ &  \vspace{0.15cm}\\
\vspace{0.4cm}\\

\hspace{0.02cm}$C=\mathrm{Extract}_R(A,M_A)$ & & \hspace{0.1cm} $S=\mathrm{Extract}_R(B,M_B)$\hspace{-0.35cm}\spa\vspace{0.1cm}\\
\hspace{3.9cm} \Big\{ \Big. \vspace{-0.7cm} & & \hspace{-0.55cm} \Big. \Big\}\\
&\hspace{-0.6cm}$\xymatrix@1@=45pt{\ar[r]^*+<4pt>{}&}$\vspace{-0.0cm}\\
\hspace{0.02cm}$c_2 \leftarrow \;\mbox{PSI-CA}(C,S)$ &\hspace{-0.6cm}$\xymatrix@1@=45pt{& \ar[l]^*+<4pt>{}_*+<4pt>{}}$ &  \vspace{0.35cm}\\
\multicolumn{3}{l}{\hspace{0.02cm}{\bf Output match as $(n-c_2)/c_1<t$}} %
\end{tabular}
\end{minipage}
}
\vspace{-0.15cm}
\caption{Privacy-preserving iris matching of biometric $A$ and $B$.}%
\label{fig:app-iris}
\end{figure}

Proposed protocol is illustrated in Fig.~\ref{fig:app-iris}. %
Given any two $n$-bit strings $X = \{x_1, \ldots, x_n\}$ and $Y = \{y_1, \ldots, y_n\}$ and a list of $k$ values $R=(r_1, \ldots, r_k)$,
with $r_i \in [1,n]$, $r_i \neq r_j$ for all $i\neq j$, we define probabilistic function $\mathrm{Extract}_{R}:\{0,1\}^n \times \{0,1\}^n \rightarrow (\{0,1\} \times \{1,\ldots,n\})^k$ as:

$$\mathrm{Extract}_{R}(X,Y) = \{w_{r_1},\ldots,w_{r_k}\}, \mbox{where }
w_{r_i} = \left\{
\begin{array}{ll}
\langle x_{r_i}, r_i \rangle & \mbox{if } y_{r_i}=1 \\
\langle r, r_i \rangle \mathrm{\ with\ } r\leftarrow\{0,1\}^\tau & \mbox{otherwise}
\end{array}
\right.\vspace{0.1cm}
$$
where $x \leftarrow Y$ represents uniform random sampling of element $x$ from set $Y$.

Intuitively, for each value $r_i$ in $R$, $\mathrm{Extract}_{R}(X,Y)$ selects the $r_i$-th bit of $X$ and encodes it with $r_i$ (e.g., concatenates the two), if the $r_i$-th bit of $Y$ is 1. If the $r_i$-th bit of $Y$ is 0, the function selects a random value end encodes it with $r_i$.

Given $A, M_A, B, M_B$, $\A$ and $\B$ privately determine $\mathrm{WHD}(A, B, M_A \wedge M_B)$:
\begin{itemize}
\item[$\bullet$] $\A$ and $\B$ negotiate $R$.
\item[$\bullet$] $\A$ computes $C_M=\mathrm{Extract}_R(M_A, M_A)$; $\B$ computes $S_M=\mathrm{Extract}_R(M_B,M_B)$.
\item[$\bullet$] $\A$ and $\B$ engage in a PSI-CA protocol where their inputs are $C_M$ and $S_M$ and  $\A$ learns the output $c_1$ of PSI-CA. $c_1$ corresponds to the number of bits set to 1 in both $M_A$ and $M_B$ at indices specified by $R$.
\item[$\bullet$] $\A$ computes $C = $ Extract$_R(A,M_A)$; similarly, $\B$ computes $S = $ Extract$_R(B,M_B)$. 
\item[$\bullet$] $\A$ and $\B$ interact in a PSI-CA protocol with input $C$ and $S$ respectively; at the end of the protocol, 
      $\A$ learns $c_2$, i.e., the size of the intersection of the subsets of $A$ and $B$ defined by $R$. 
\item[$\bullet$] Biometric $A$ matches $B$ iff $(n-c_2)/c_1<t$.
\end{itemize}

\subsection{Comparison to prior work}
We now compare our technique for privacy-preserving iris matching to prior work, namely the 
technique in~\cite{Blanton11esorics}. We compare our approach with the protocol 
in~\cite{Blanton11esorics} because, at the time of writing, it provides the best performance for privacy-preserving comparison of iris codes.
First, observe that protocol in Fig.~\ref{fig:app-iris} estimates the Weighted Hamming Distance
with bounded error, whereas, construction in~\cite{Blanton11esorics} yields its exact computation.
However, as we discuss below, the error incurred by our technique is low enough to be used in practice and achieves reduced computational complexity.
In fact, our probabilistic protocol could be used to perform a fast, preliminary test: if differences between two irises are significant, then there is no need for further tests. Otherwise, the two parties can engage in the protocol in \cite{Blanton11esorics} to obtain (in a privacy-preserving way) a precise result. %
Next, as opposed to the technique in \cite{Blanton11esorics}, $\A$ also learns an estimate on the number of bits set to 1 in the combined mask $M_A \wedge M_B$, but not their position. 
However, this information is not sensitive, thus, it does not leak any information about the iris sampled by $\A$ or $\B$.

\paragraph{Optimization.} As discussed above, it is reasonable to assume that $\B$ (e.g., the Department of Homeland Security) holds a database with a large number of biometric samples, whereas, $\A$ (e.g., the Transportation Security Administration) has only one or few samples that she is searching
in  $\B$'s database. To this end, we now show how the protocol in Fig.~\ref{fig:app-iris} can be optimized,
by pre-computing several expensive operations offline, for such a scenario.

Note that $\B$ can perform the offline phase of \DGT\ (see Fig.~\ref{fig:psi-ca}) on all bits 
of his biometric samples: unlike the protocol in~\cite{Blanton11esorics}, this is required {\em 
only once}, independently on the number of interactions between $\B$ and any user.

After negotiating with $\B$ the values $R=(r_1, \ldots, r_k)$, and before receiving her input, $\A$ pre-computes $k$ pairs $\langle\alpha_{0,i}=H(\langle 0,r_i\rangle)^{R'_c}, \alpha_{1,i}= H(\langle 1,r_i\rangle)^{R'_c}\rangle$. (This assumes the use of \DGT.) 
Once $\A$'s mask has been revealed, she constructs the corresponding encrypted representation by simply selecting the appropriate element from each pair.
Similarly, she computes $k$ triples $(\alpha_{0,i}, \alpha_{1,i}, \alpha_{\rho,i})$ where $\alpha_{0,i}, \alpha_{1,i}, \alpha_{\rho,i}$ represent 0, 1 and a random element in ${\{0,1\}^\tau}$, respectively. $\A$ later uses such triples to represent each bit $\beta_i$ of her iris sample as $\alpha_{i}=\alpha_{\beta,i}$ if the corresponding bit in the mask is 1, else, as $\alpha_{i}=\alpha_{\rho, i}$.
During the online phase, $\A$ selects the appropriate pre-computed values to match the  mask and the iris bits. Similarly, $\B$ inputs the selected bits of each record's mask and iris into the PSI-CA protocol.

\paragraph{Performance Comparison.}
In Table~\ref{tab:perf}, we report running times from implementations of, respectively, our protocol in 
Fig.~\ref{fig:app-iris} and technique in~\cite{Blanton11esorics}.
We assume that about 75\% of the bits in the mask are set to 1 (like 
in~\cite{Blanton11esorics}). We set the length of each iris and mask to 2048 bits and the database size to 320 irises, which is the number used in prior work. All tests are run on a single Intel Xeon E5420 core running at 2.50GHz. We set $k=25$, thus, obtaining an expected error in the order of $1/\sqrt{25}$, i.e., 20\%.

\begin{table}[t]
\vspace{-0.2cm}
{ 
\setlength{\tabcolsep}{0.5ex}
\centerline{\resizebox{0.8\columnwidth}{!}{\small
\begin{tabular}{|c|c|c|c|} \hline
\multicolumn{4}{|c|}{\bf \em Protocol in Fig.~\ref{fig:app-iris}}\\ \hline
\multicolumn{2}{|c|}{} & {\bf Offline} & {\bf Online} \\ \hline
$\B$ & $\pm$ 5-bit rot. & 0.13 ms + 5.8 s/rec & 71.5 ms/rec \\ \cline{2-4}
& no rot. & 0.13 ms + 530 ms/rec & 6.5 ms/rec \\\cline{1-4}
$\A$ & $\pm$ 5-bit rot. & 71.63 ms & 71.5 ms/rec \\ \cline{2-4} 
& no rot. & 6.63 ms &  6.5 ms/rec \\ 
\hline
\end{tabular}
\begin{tabular}{|c|c|} \hline
\multicolumn{2}{|c|}{{\bf \em Protocol in }\cite{Blanton11esorics}}\\ \hline
{\bf Offline} & {\bf Online} \\ \hline
2.8 s + 71.55 ms/rec & 97.2 ms + 134.28 ms/rec \\ \hline
2.6 s + 6.48 ms/rec & 97.2 ms + 12.33 ms/rec \\\hline
12.2 s + 3 ms/rec & 20.34 ms/rec \\ \hline 
11.7 s + 0.27 ms/rec &  1.8 ms/rec \\ 
\hline
\end{tabular}}}
\vspace{-0.25cm}
\caption{Computation overhead of our randomized iris matching technique in Fig.~\ref{fig:app-iris}
and that of \cite{Blanton11esorics}. Experiments are performed with 5-bit left/right rotation and with no rotation of the iris sample. ``rot'' abbreviates ``rotation'' and ``rec'' -- ``record''.}
\label{tab:perf}
}
\vspace{-0.05cm}
\end{table}

Observe that {\em online} cost incurred by $\B$ with our technique is significantly lower compared to that of protocol in~\cite{Blanton11esorics}. Whereas, it is higher for $\A$. Nonetheless, summing up the computation overhead incurred by both $\A$ and $\B$, our protocol always results faster that the one in~\cite{Blanton11esorics} for the online computation. 

The {\em offline} cost imposed on $\B$ is about twice as high as its counterpart in protocol from \cite{Blanton11esorics}.  However, in our protocol, the offline part is done once, for all possible interactions, independently from their number. Whereas, in \cite{Blanton11esorics}, the offline computation needs to be performed anew, for {\em every} interaction. In settings where $\B$ interacts frequently with multiple entities, this may significantly effect protocol's overall efficiency. Furthermore, the offline cost imposed on $\A$ is markedly lower (several orders of magnitudes) using our technique.

We conclude that the protocol in Fig.~\ref{fig:app-iris} improves, in many settings, overall efficiency compared to state of the art. However, it introduces a maximum error of about 20\%, whereas, the scheme in \cite{Blanton11esorics} compute the exact -- rather than approximate -- outcome of an iris comparison. 
Thus, a good practice is to use the scheme in Fig.~\ref{fig:app-iris} to perform an initial selection of relevant biometric samples, using a threshold $t'>t$, in order to compensate for the error. The final matching on selected samples can then be done, in a privacy-preserving manner, using the protocol in \cite{Blanton11esorics}.

\section{Privacy-Preserving Multimedia File Similarity}\label{sec:multimedia}
Amid widespread availability of digital cameras, digital audio recorders, and media-enabled smartphones, users generate a staggering amount of {\em multimedia} content.
As a result, secure online storage (and management) of large volumes of multimedia data
becomes increasingly desirable. %
According to \cite{youtubeupload11}, YouTube received more than 13 million hours of video in 2010, and
48 hours of new content are uploaded {\em every minute} (i.e., 8 years of video each day).
Similarly, Flickr users upload about 60 photos every second.

Such an enormous amount of multimedia data prevents manual content curation -- e.g., manually 
{\em tagging} all uploaded content to allow textual searches. 
For this reason, in recent years research has focused on automated tools for feature 
extraction and analysis on multimedia content. 
A prominent example is Content-Based Image Retrieval (CBIR)~\cite{smeulders2000content}. It 
allows automatic extraction of features from an image, a video, an audio file or any other 
multimedia content. These features can then be compared across different files, establishing for example {\em how similar} two documents are.
There are several available techniques to implement CBIR,
including search techniques based on color histograms~\cite{smith96}, 
bin similarity coefficients \cite{niblack93}, texture for image characterization \cite{ma96},
shape features \cite{sclaroff97}, edge directions \cite{jain96}, and matching of shape components such as corners, line segments or circular arcs \cite{cohen97}.

In this section we design a generic privacy-preserving technique for comparing multimedia 
documents by comparing their features. Our technique is based on Jaccard and MinHash, and is 
oblivious to the specific type of features used to perform comparison.We implement a small 
prototype, which we use to evaluate the performance of our approach.

\paragraph{Prior Work.} 
Motivated by the potential sensitivity of multimedia data, the research community has begun to
develop mechanisms for secure signal processing.
For instance, authors in~\cite{erkin2007} are the first to investigate secure signal processing related to
multimedia documents. 
Then, the work in~\cite{lu09-2,lu09} introduces two protocols to search over encrypted multimedia databases.
Specifically, it extracts 256 visual features from each image. Then, files are encrypted in a distance-preserving fashion,
so that encrypted features can be directly compared for similarity evaluation. Similarity is computed using the Jaccard index 
between the visual features of searched image and those of images in a database.
However, the security of the scheme relies on order-preserving encryption (used to mask frequencies of recurring visual features), which is known to provide only a limited level of security \cite{boldyreva-ope}.

\begin{figure}[h]
\centering
\hspace{1cm}
\fbox{\small
\begin{minipage}{0.61\columnwidth}
\begin{tabular}{lcl}
\hspace{-0.05cm}$\A$ $(F_A)$ & &\hspace{0.3cm} $\B$ $(F_{B:i})$ \vspace{-0.2cm}\\
\multicolumn{3}{c}{\hspace{-0.3cm}\line(1,0){293.8} \vspace{0.1cm}}\\
\hspace{-0.05cm}$A\leftarrow\mbox{Extract}(F_A)$ & & \hspace{0.3cm} $B_i\leftarrow\mbox{Extract}(F_{B:i})$\hspace{0.2cm}\\
\hspace{3.99cm} \Big\{ \Big. \vspace{-0.69cm} & & \hspace{-0.55cm} \Big. \Big\}\\
&\hspace{-0.62cm}$\xymatrix@1@=45pt{\ar[r]^*+<4pt>{}&}$\vspace{0.02cm}\\
 &\hspace{-0.62cm}$\xymatrix@1@=45pt{& \ar[l]^* +<4pt>{}_*+<4pt>{}}$ &  \vspace{-0.2cm}\\
\\
\hspace{-0.05cm}$|A\cap B_i| \leftarrow \mbox{PSI-CA}(A,B_i)$\vspace{-0.05cm}\\
\end{tabular}
\vspace{0.2cm}

\hspace{0.14cm}{\bf Output Similarity as} %
$J(A,B_i)=\dfrac{|A\cap B_i|}{|A|+|B_i|-|A\cap B_i|}$ %
\end{minipage}
}
\caption{Privacy-preserving  evaluation of multimedia file similarity.}
\label{fig:ppmfs}
\end{figure}
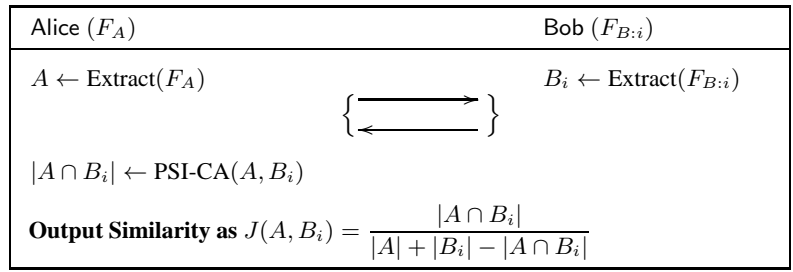

\paragraph{Our Approach.}
We use the Jaccard index to assess the similarity of multimedia files.
As showed in Section~\ref{sec:new}, we can do so, in a privacy-preserving way,
using protocol in Fig.~\ref{fig:sjacc}. Our Privacy-preserving evaluation of multimedia file similarity protocol is presented in Fig.~\ref{fig:ppmfs}. We denote a multimedia file owned by \A\ as $F_A$, and a file owned by Bob as $F_{B:i}$.

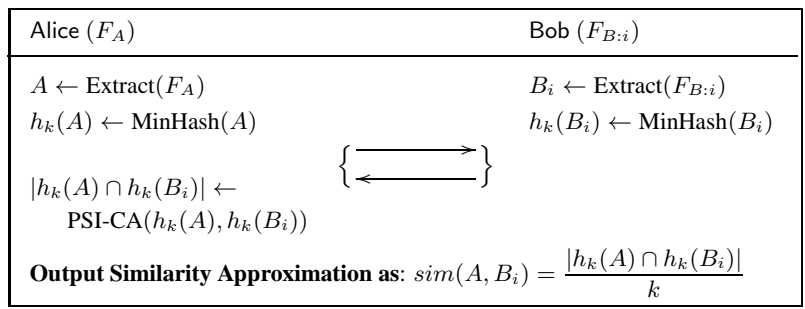
\begin{figure}[h]
\centering
\hspace{1cm}
\fbox{\small
\begin{minipage}{0.62\columnwidth}
\begin{tabular}{lcl}
\hspace{-0.05cm}$\A$ $(F_A)$ & &\hspace{-3.8cm} $\B$ $(F_{B:i})$ \vspace{-0.15cm}\\
\multicolumn{3}{c}{\hspace{-0.5cm}\line(1,0){299} \vspace{0.1cm}}\\
\hspace{-0.05cm}$A\leftarrow\mbox{Extract}(F_A)$ & & \hspace{-3.8cm} $B_i\leftarrow\mbox{Extract}(F_{B:i})$\vspace{0.1cm}\\
\hspace{-0.05cm}$h_k(A)\leftarrow\mbox{MinHash}(A)$ & & \hspace{-3.8cm} $h_k(B_i)\leftarrow\mbox{MinHash}(B_i)$\hspace{-0.35cm}\spa\vspace{0.1cm}\\
\hspace{3.95cm} \Big\{ \Big. \vspace{-0.7cm} & & \hspace{-4.55cm} \Big. \Big\}\\
&\hspace{-9.6cm}$\xymatrix@1@=45pt{\ar[r]^*+<4pt>{}&}$\vspace{-0.0cm}\\
 &\hspace{-9.6cm}$\xymatrix@1@=45pt{& \ar[l]^*+<4pt>{}_*+<4pt>{}}$ &  \vspace{-0.25cm}\\
\hspace{-0.05cm}$|h_k(A)\cap h_k(B_i)|\leftarrow$\vspace{0.05cm}\\
$\;\;\;\;\;\mbox{PSI-CA}(h_k(A),h_k(B_i))$\vspace{0.15cm}\\
\hspace{-0.05cm}{\bf Output Similarity Approximation as}: %
$sim(A,B_i)=\dfrac{|h_k(A)\cap h_k(B_i)|}{k}$ %
\end{tabular}
\end{minipage}
}
\vspace{-0.2cm}
\caption{Privacy-preserving {\em approximation} of multimedia file similarity.}
\label{fig:ppmfs-mh}
\vspace{-0.2cm}
\end{figure}

Our approach is independent of the underlying feature extraction algorithm, even though protocol accuracy naturally relies on the quality of the feature extraction phase. 
Once features have been extracted, our privacy-preserving protocols only reveal their similarity, thus,
without disclosing the features themselves.
As an example, we instantiate our techniques for privacy-preserving {\em image} similarity.
Our approach for feature extraction is based on \cite{lu09}, since its accuracy is reasonable enough for
real-world use, using color histograms in the color space of Hue, Saturation and Value (HSV).
Thus, our scheme achieves the same accuracy of~\cite{lu09}, in terms of precision and recall.
Once again, to obtain improved efficiency, similarity can be approximated using MinHash techniques,
as per protocol in Fig.~\ref{fig:ppmfs-mh}. In this case, protocol performance and accuracy depend on the MinHash parameter $k$.

\paragraph{Performance Evaluation.} We test our technique with the same dataset used by \cite{lu09}, i.e., 1000 images from the standard Corel dataset. 
We extract 256 features from each image, for a total of 256,000 features for the whole database.
We envision a user, $\A$, willing to assess similarity of an image against an image database, held by $\B$.
 We run our protocol for privately computing the Jaccard index (``Exact'' rows in Table~\ref{tab:multimedia}) and for estimated similarity, using MinHash with $k=100$ (``Approximate'' row). Table~\ref{tab:multimedia} summarizes our experiments. All tests are run on a single Intel Xeon E5420 core running at 2.50GHz and show that privacy protection
is attainable at a very limited cost.

Remark that a thorough performance comparison between our protocol and related work is out of the scope of this paper, since the main effort of prior work has been achieving high accuracy in similarity detection, rather improving efficiency.
Thus, we defer it to future work. Nonetheless, the authors of~\cite{lu09} report
that the running time of their protocol is in the order of 1 second per image,
on a hardware comparable to our testbed (a dual-core 3GHz PC with 4GB of RAM).
Therefore, it is safe to assume that our protocol for privacy-preserving multimedia file similarity is about one order of magnitude faster than available techniques, even without considering pre-computation. 
\begin{table}[t]
{ 
\setlength{\tabcolsep}{1ex}
\centerline{{\small
\begin{tabular}{|c|c|c|c|} \hline
& & {\bf Offline} & {\bf Online} \\ \hline
$\B$ & Exact & 0.13 ms + 33.28 ms/record & 33.28 ms/record \\ \cline{2-4}
& Approximate & 0.13 ms + 13 ms/record & 13 ms/rec \\\cline{1-4}
$\A$ & Exact &  66.69 ms & 33.28 ms/record \\ \cline{2-4} 
& Approximate & 26.13 ms &  13 ms/record \\ 
\hline
\end{tabular}}}
\vspace{-0.3cm}
\caption{Computation cost of our multimedia documents similarity protocol.}
\label{tab:multimedia}
}
\vspace{-0.05cm}
\end{table}

\section{Faster and Size-Hiding (Approximated) PSI-CA}
\label{sec:approx-psica}
Privacy-preserving computation of set intersection cardinality has been investigated
quite extensively by the research community~\cite{ePrint,freedman2004efficient,kissner2005,vaidya2005},
motivated by several interesting applications, including: 
privacy-preserving authentication and key exchange protocols~\cite{ateniese2007secret}, 
data and association rule mining~\cite{vaidya2005},
genomic applications~\cite{baldi2011countering}, healthcare~\cite{kissner2005},
policy-based information sharing~\cite{ePrint}, anonymous routing~\cite{zhang2005anonymous}, 
and -- as argued by this paper -- sample set similarity.

In many of the application scenarios, however, it may be enough to obtain  an estimation, rather than the exact measure,
of set intersection cardinality. For instance, if PSI-CA is used to privately quantify the
number of common social-network friends (e.g., to assess profile similarity)~\cite{li2011findu}, then
one may want to trade off a bounded accuracy loss with 
a significant improvement in protocol overhead (and without sacrificing the level of attained privacy protection).
Clearly, such an improved construction is particularly appealing whenever participants' input sets are very large.

Using MinHash techniques, we can construct a protocol for privacy-preserving estimation of set 
intersection cardinality with (constant) computation and communication
complexities, that only depend on the MinHash parameter -- i.e., $O(k)$.
Proposed construction is illustrated in Fig.~\ref{fig:csi}.
While we tolerate a bounded accuracy loss depending on MinHash's parameter 
$k$, i.e., $O(1/\sqrt{k})$, observe that our protocol achieves the same, provably-secure, privacy guarantees as if we ran PSI-CA
on whole sets.

\begin{figure}[htb]
\vspace{0.2cm}
\begin{center}
\fbox{\small
\begin{minipage}{0.72\columnwidth}
\underline{\textbf{Privacy-preserving approximation of $|A\cap B|$}}\vspace{0.05cm}\\
{\small {\sf Run by $\A$ and $\B$, on input, resp., $A$ and $B$}}\vspace{-0.1cm}
\begin{enumerate}
\item  $\A$ and $\B$  compute, $\{\langle a_i,i\rangle\}_{i=1}^k$ and
$\{\langle b_i,i\rangle\}_{i=1}^k$, resp.,  using multi-hash\\ MinHash
where: $a_i\stackrel{\text{\tiny def}}{=}h^{(i)}_{min}(A)$ and $b_i\stackrel{\text{\tiny def}}{=}h^{(i)}_{min}(B)$
\item $\A$ and $\B$ execute PSI-CA on input, resp., $(\{\langle a_i,i\rangle\}_{i=1}^k, k)$ and
$(\{\langle b_i,i\rangle\}_{i=1}^k,k)$
\item $\A$ learns $\delta=|\{\langle a_i,i\rangle\}_{i=1}^k\cap\{\langle b_i,i\rangle\}_{i=1}^k|$
\item $\B$ sends $w$ to $\A$	
\item $\A$ outputs $\delta \cdot (v+w)/(1+\delta)$
\end{enumerate}
\end{minipage}}
\vspace{-0.25cm}
\caption{Our technique for Approximated Private Set Intersection Cardinality.}
\label{fig:csi}
\end{center}
\vspace{-0.35cm}
\end{figure}

\paragraph{Size-Hiding.} Another factor motivating the use of MinHash techniques for PSI-CA is related to input size secrecy.
Available PSI-CA protocols always disclose, from the execution, at least an upper bound on input 
set sizes. Whereas, protocol in Fig.~\ref{fig:csi} conceals -- unconditionally -- $\A$'s set size, thus, 
achieving {\em Size-Hiding} Private Set Intersection Cardinality. With this protocol, \A\ and \B\ do not need to disclose $|A|$ and $|B|$. Rather, public protocol input includes $k$, which is {\bf\em independent from $|A|$ and $|B|$}. Because secure PSI-CA, used as building block for the protocol in Fig.~\ref{fig:csi}, does not leak information about the input sets, neither party learns information about the ratio between $k$ and $|A|$, $|B|$.
Considering recent results motivating 
the need for size-hiding features in private set operations (see~\cite{ateniese2011if}), 
this additional feature is particularly valuable.

\paragraph{Note:} While we leave as part of future work a thorough experimental performance evaluation, observe that PSI-CA and the approximated and size-hiding protocol (using MinHash) are essentially the same protocols. The latter runs on smaller, constant-size input ($k$): since the protocols have linear complexities, it is straightforward to assess the performance spread. 

\section{Conclusion}\label{sec:conclusion}
This paper introduced the first efficient construction for privacy-preserving evaluation
of sample set similarity, relying on the Jaccard index measure.
We also presented an efficient randomized protocol that approximates,
with bounded error, this similarity index.
Our techniques are generic and practical enough to be used as a basic building block for a wide
array of different privacy-preserving functionalities, including 
document and multimedia file similarity, biometric matching, genomic testing, similarity of 
social profiles, and so on.
Experimental analyses support our efficiency claims and demonstrate improvements over prior results.
As part of future work, we plan to study privacy-preserving computation of other similarity measures,
as well as to further investigate additional applications and extensions.

\descr{Acknowledgments.} The work of Carlo Blundo has been supported, in part, by 
the Italian Ministry of Research (MIUR), under project n.~2010RTFWBH ``Data-Driven Genomic Computing (GenData 2020).''
We gratefully acknowledge Elaine Shi 
for providing us with the idea of genetic paternity testing 
as one of our motivating examples. Finally, we thank the Journal of Computer Security's (anonymous) reviewers,
whose comments helped us improve papers' content and presentation.

{\small
\bibliographystyle{abbrv}
\bibliography{bibfile}

\begin{thebibliography}{10}

\bibitem{AtallahBLFT04}
M.~J. Atallah, M.~Bykova, J.~Li, K.~B. Frikken, and M.~Topkara.
\newblock Private collaborative forecasting and benchmarking.
\newblock In {\em WPES}, 2004.

\bibitem{ateniese2007secret}
G.~Ateniese, M.~Blanton, and J.~Kirsch.
\newblock Secret handshakes with dynamic and fuzzy matching.
\newblock In {\em NDSS}, 2007.

\bibitem{ateniese2011if}
G.~Ateniese, E.~De~Cristofaro, and G.~Tsudik.
\newblock {(If) size matters: Size-Hiding Private Set Intersection}.
\newblock In {\em PKC}, 2011.

\bibitem{pino}
E.~Baglioni, L.~Becchetti, L.~Bergamini, U.~Colesanti, L.~Filipponi,
  A.~Vitaletti, and G.~Persiano.
\newblock A lightweight privacy preserving sms-based recommendation system for
  mobile users.
\newblock In {\em RecSys}, 2010.

\bibitem{baldi2011countering}
P.~Baldi, R.~Baronio, E.~De~Cristofaro, P.~Gasti, and G.~Tsudik.
\newblock Countering gattaca: efficient and secure testing of fully-sequenced
  human genomes.
\newblock In {\em CCS}, 2011.

\bibitem{bar10}
M.~Barni, T.~Bianchi, D.~Catalano, M.~{Di Raimondo}, R.~Labati, P.~Failla,
  D.~Fiore, R.~Lazzeretti, V.~Piuri, F.~Scotti, and A.~Piva.
\newblock Privacy-preserving fingercode authentication.
\newblock In {\em MM\&Sec}, 2010.

\bibitem{Blanton11esorics}
M.~Blanton and P.~Gasti.
\newblock {Secure and Efficient Protocols for Iris and Fingerprint
  Identification}.
\newblock In {\em ESORICS}, 2011.

\bibitem{boldyreva-ope}
A.~Boldyreva, N.~Chenette, Y.~Lee, and A.~O'Neill.
\newblock Order-preserving symmetric encryption.
\newblock In {\em EUROCRYPT}, 2009.

\bibitem{broder1997resemblance}
A.~Broder.
\newblock On the resemblance and containment of documents.
\newblock In {\em Compression and Complexity of Sequences}, 1997.

\bibitem{broder1998min}
A.~Broder, M.~Charikar, A.~Frieze, and M.~Mitzenmacher.
\newblock Min-wise independent permutations.
\newblock In {\em STOC}, 1998.

\bibitem{bunn2007secure}
P.~Bunn and R.~Ostrovsky.
\newblock Secure two-party k-means clustering.
\newblock In {\em CCS}, 2007.

\bibitem{chernoff1952measure}
H.~Chernoff.
\newblock A measure of asymptotic efficiency for tests of a hypothesis based on
  the sum of observations.
\newblock {\em The Annals of Mathematical Statistics}, 1952.

\bibitem{CimatoGPSS09}
S.~Cimato, M.~Gamassi, V.~Piuri, R.~Sassi, and F.~Scotti.
\newblock {\em Privacy in Biometrics}.
\newblock Wiley-IEEE Press, 2009.

\bibitem{cohen97}
S.~D. Cohen and L.~J. Guibas.
\newblock Shape-based image retrieval using geometric hashing.
\newblock In {\em ARPA Image Understanding Workshop}, 1997.

\bibitem{kdd}
{Cornell Univ.}
\newblock {KDDCUP Dataset}.
\newblock \url{http://www.cs.cornell.edu/projects/kddcup/datasets.html}.

\bibitem{genodroid}
E.~D. Cristofaro, S.~Faber, P.~Gasti, and G.~Tsudik.
\newblock Genodroid: are privacy-preserving genomic tests ready for prime time?
\newblock In {\em WPES}, 2012.

\bibitem{ePrint}
E.~{De Cristofaro}, P.~Gasti, and G.~Tsudik.
\newblock {Fast and Private Computation of Cardinality of Set Intersection and
  Union}.
\newblock In {\em CANS}, 2012.

\bibitem{dombek2000use}
P.~Dombek, L.~Johnson, S.~Zimmerley, and M.~Sadowsky.
\newblock {Use of repetitive DNA sequences and the PCR to differentiate
  Escherichia coli isolates from human and animal sources}.
\newblock {\em Applied and Environmental Microbiology}, 66(6), 2000.

\bibitem{erk09}
Z.~Erkin, M.~Franz, J.~Guajardo, S.~Katzenbeisser, I.~Lagendijk, and T.~Toft.
\newblock Privacy-preserving face recognition.
\newblock In {\em PETS}, 2009.

\bibitem{erkin2007}
Z.~Erkin, A.~Piva, S.~Katzenbeisser, R.~L. Lagendijk, J.~Shokrollahi, G.~Neven,
  and M.~Barni.
\newblock Protection and retrieval of encrypted multimedia content: When
  cryptography meets signal processing.
\newblock {\em EURASIP J. Information Security}, 2007, 2007.

\bibitem{freedman2004efficient}
M.~Freedman, K.~Nissim, and B.~Pinkas.
\newblock Efficient private matching and set intersection.
\newblock In {\em EUROCRYPT}, 2004.

\bibitem{Goldreich}
O.~Goldreich.
\newblock {\em {Foundations of cryptography}}.
\newblock Cambridge Univ Press, 2004.

\bibitem{uaei}
{IrisGuard Press Release}.
\newblock {\url{http://cl.ly/3KIB}}.

\bibitem{jaccard1901etude}
P.~Jaccard.
\newblock {Etude comparative de la distribution florale dans une portion des
  Alpes et du Jura}, 1901.

\bibitem{jai00}
A.~Jain, S.~Prabhakar, L.~Hong, and S.~Pankanti.
\newblock Filterbank-based fingerprint matching.
\newblock {\em IEEE Transactions on Image Processing}, 9(5), 2000.

\bibitem{jain96}
A.~K. Jain and A.~Vailaya.
\newblock Image retrieval using color and shape.
\newblock {\em Pattern Recognition}, 1996.

\bibitem{jiang2008similar}
W.~Jiang, M.~Murugesan, C.~Clifton, and L.~Si.
\newblock Similar document detection with limited information disclosure.
\newblock In {\em ICDE}, 2008.

\bibitem{JiSa}
W.~Jiang and B.~K. Samanthula.
\newblock N-gram based secure similar document detection.
\newblock In {\em DBSec 2011}, Lecture Notes in Computer Science. Springer,
  2011.

\bibitem{rulemining}
M.~Kantarcioglu, R.~Nix, and J.~Vaidya.
\newblock {An efficient approximate protocol for privacy-preserving association
  rule mining}.
\newblock In {\em KDD}, 2009.

\bibitem{katti2005collaborating}
S.~Katti, B.~Krishnamurthy, and D.~Katabi.
\newblock Collaborating against common enemies.
\newblock In {\em IMC}, 2005.

\bibitem{kerschbaum2012outsourced}
F.~Kerschbaum.
\newblock {Outsourced Private Set Intersection Using Homomorphic Encryption}.
\newblock In {\em AsiaCCS}, 2012.

\bibitem{kissner2005}
L.~Kissner and D.~Song.
\newblock {Privacy-preserving set operations}.
\newblock In {\em CRYPTO}, 2005.

\bibitem{li2011findu}
M.~Li, N.~Cao, S.~Yu, and W.~Lou.
\newblock {Findu: Privacy-preserving personal profile matching in mobile social
  networks}.
\newblock In {\em INFOCOM}, 2011.

\bibitem{lipmaa2003verifiable}
H.~Lipmaa.
\newblock {Verifiable Homomorphic Oblivious Transfer and Private Equality
  Test}.
\newblock In {\em ASIACRYPT}, 2003.

\bibitem{lu09-2}
W.~Lu, A.~Swaminathan, A.~L. Varna, and M.~Wu.
\newblock Enabling search over encrypted multimedia databases.
\newblock In {\em Media Forensics and Security}, 2009.

\bibitem{lu09}
W.~Lu, A.~L. Varna, A.~Swaminathan, and M.~Wu.
\newblock Secure image retrieval through feature protection.
\newblock In {\em ICASSP}, 2009.

\bibitem{ma96}
W.-Y. Ma and B.~S. Manjunath.
\newblock Texture features and learning similarity.
\newblock In {\em CVPR}, 1996.

\bibitem{manber1994finding}
U.~Manber.
\newblock Finding similar files in a large file system.
\newblock In {\em USENIX}, 1994.

\bibitem{MurugesanVLDB2010}
M.~Murugesan, W.~Jiang, C.~Clifton, L.~Si, and J.~Vaidya.
\newblock Efficient privacy-preserving similar document detection.
\newblock {\em The VLDB Journal}, 19, August 2010.

\bibitem{niblack93}
W.~Niblack, R.~Barber, W.~Equitz, M.~Flickner, E.~H. Glasman, D.~Petkovic,
  P.~Yanker, C.~Faloutsos, and G.~Taubin.
\newblock {The QBIC Project: Querying Images by Content, Using Color, Texture,
  and Shape}.
\newblock In {\em SPIE}, 1993.

\bibitem{49}
{Ookla Net Metrics}.
\newblock {Canada and US Source Data}.
\newblock \url{http://www.netindex.com/source-data/}, 2011.

\bibitem{scifi}
M.~Osadchy, B.~Pinkas, A.~Jarrous, and B.~Moskovich.
\newblock {SCiFI} -- {A} system for secure face identification.
\newblock In {\em IEEE S\&P}, 2010.

\bibitem{Pa99}
P.~Paillier.
\newblock Public-key cryptosystems based on composite degree residuosity
  classes.
\newblock In {\em EUROCRYPT}, 1999.

\bibitem{popescu2006fuzzy}
M.~Popescu, J.~Keller, and J.~Mitchell.
\newblock Fuzzy measures on the gene ontology for gene product similarity.
\newblock {\em Transactions on Computational Biology and Bioinformatics}, 2006.

\bibitem{PSDM04}
P.~Ravikumar, W.~Cohen, and S.~Fienberg.
\newblock A secure protocol for computing string distance metrics.
\newblock In {\em PSDM}, 2004.

\bibitem{sad09}
A.-R. Sadeghi, T.~Schneider, and I.~Wehrenberg.
\newblock Efficient privacy-preserving face recognition.
\newblock In {\em ICISC}, 2009.

\bibitem{sclaroff97}
S.~Sclaroff.
\newblock Deformable prototypes for encoding shape categories in image
  databases.
\newblock {\em Pattern Recognition}, 30(4), 1997.

\bibitem{singh2009privacy}
M.~Singh, P.~Krishna, and A.~Saxena.
\newblock A privacy preserving jaccard similarity function for mining encrypted
  data.
\newblock In {\em TENCON}, 2009.

\bibitem{smeulders2000content}
A.~Smeulders, M.~Worring, S.~Santini, A.~Gupta, and R.~Jain.
\newblock Content-based image retrieval at the end of the early years.
\newblock {\em IEEE Transactions on Pattern Analysis and Machine Intelligence},
  22(12), 2000.

\bibitem{smith96}
J.~R. Smith and S.-F. Chang.
\newblock Tools and techniques for color image retrieval.
\newblock In {\em SPIE}, 1996.

\bibitem{strehl2000impact}
A.~Strehl, J.~Ghosh, and R.~Mooney.
\newblock Impact of similarity measures on web-page clustering.
\newblock In {\em AAAI}, 2000.

\bibitem{tan2006introduction}
P.~Tan, M.~Steinbach, V.~Kumar, et~al.
\newblock {\em Introduction to data mining}.
\newblock Pearson, 2006.

\bibitem{usvisit}
{U.S. Department of Homeland Security}.
\newblock {\url{http://www.dhs.gov/files/programs/usv.shtm}}.

\bibitem{vaidya2005}
J.~Vaidya and C.~Clifton.
\newblock {Secure set intersection cardinality with application to association
  rule mining}.
\newblock {\em Journal of Computer Security}, 13(4), 2005.

\bibitem{xiao2008efficient}
C.~Xiao, W.~Wang, X.~Lin, and J.~Yu.
\newblock Efficient similarity joins for near duplicate detection.
\newblock In {\em WWW}, 2008.

\bibitem{yao86}
A.~Yao.
\newblock How to generate and exchange secrets.
\newblock In {\em FOCS}, 1986.

\bibitem{youtubeupload11}
{Youtube press release}.
\newblock {\url{http://www.youtube.com/t/press\_statistics}}.

\bibitem{zhang2005anonymous}
Y.~Zhang, W.~Liu, and W.~Lou.
\newblock Anonymous communications in mobile ad hoc networks.
\newblock In {\em INFOCOM}, volume~3, 2005.

\end{thebibliography}
}

\appendix

\section{Additional Details on MinHash Techniques}\label{app:minhash}

\paragraph{Single-Hash MinHashes.} Besides the multiple-hash technique presented in Section~\ref{sec:preliminaries},
another approach for approximating the Jaccard index using MinHash employs a single hash function.
In this case, rather than selecting one value per hash function, one selects the $k$ values from set $A$ that hash the to smallest integers.
Specifically, let $h(\cdot)$ be a hash function, and $k$ a fixed parameter;
for any set $S$, define $h_k(S) \subset S$ as the set of the pre-images of the $k$ smallest hash values of elements of $S$.
Consider: \vspace{-0.05cm} %
\begin{equation}%
sim(A,B) = \frac{\left|(h_k(h_k(A)\cup h_k(B))) \cap (h_k(A)\cap h_k(B)) \right|}{|h_k(h_k(A) \cup h_k(B))|}\vspace{-0.05cm}
\end{equation}
It holds that $sim(A,B)$ is an unbiased estimate of the Jaccard index of $A$ and $B$.
Again, by standard Chernoff bounds~\cite{chernoff1952measure}, the expected error is $O(1/\sqrt{k})$.

\paragraph{MinHash Instantiations.} 
In order to implement the MinHash schemes described in this paper, the hash function should be 
defined by a random permutation over the set $A\cup B$. Assuming $m=|A\cup B|$, then one
would need $\Omega(m\log{m})$ bits to specify a truly random permutation, thus, yielding an infeasible overhead even for small values of $m$.
Broder, et al.~\cite{broder1998min} point out that one can obviate
to this problem by using {\em Min-wise Independent Permutation} (MWIP) families rather than random permutations.
Using MWIPs, for any subset of the domain, any element is equally likely to be the minimum,
but the number of bits to specify such a permutation is showed to be still be relatively large, i.e.,  $\Omega(m)$.
In practice, however, one can allow certain relaxations. To this end, \cite{broder1998min} introduces
{\em approximate} MWIPs, by accepting a small error $\varepsilon$.
The authors require all items in a set $S$ to have only a (almost equal) chance to become the minimum element
of $A$'s image under the permutation. Thus, for any approximate MWIP, implemented using $h^{(i)}_{min}(\cdot)$ as defined above, 
for an expected relative error $\varepsilon$,
it holds:\vspace{-0.05cm}
$$\left|\Pr\left[h^{(i)}_{min}(S)= s\right]-\dfrac{1}{|S|}\right|\leq\dfrac{\varepsilon}{|S|}\vspace{-0.05cm}$$
A class of permutations often used in practice is one based on {\em linear transformations}. It is 
assumed that the universe is $\Z_p$ for some prime $p$ and the family of permutation is constructed 
using a hash function computed as $h(x) = ax + b \mod p$, where $(a, b) \in (\Z_p^*, \Z_p)$. 
Such a linear transformations are easy to represent and efficiently calculable. 

\section{Flaw in Private Document Similarity in~\cite{JiSa}}\label{app:flaw}

In this section, we show that the protocol in \cite{JiSa} is not privacy-preserving (even in semi-honest model). 
In fact, $\B$, in order to participate in the protocol, must disclose his global space of tri-grams. Given this information, $\A$ can efficiently check whether a word, e.g., $w$, appears in $\B$'s document 
collection. Indeed, $\A$ computes $w$'s tri-gram based representation, then she checks whether all such 
tri-grams appear in $\B$'s public global space. If so, $\A$ learns that $w$ appears in a document held by $\B$ with some non-zero probability. Technically, this probability is not 1 because $\A$ could have a {\em false positive}, i.e. $w$ may not be in $\B$'s documents even though $w$'s trigrams are in $\B$'s public global space. On the other hand, if at least one of the tri-grams of $w$ is not in $\B$'s public global space, $\A$ learns that $\B$'s documents do not contain $w$. This, obviously, violates privacy requirements.
If $\A$ and $\B$ include punctation and spaces in their tri-grams representation of their documents, the probability of \emph{false positive} becomes negligible. 
We do not exploit ``relations'' between consecutive meaningful words in the sentence, 
which could potentially (further) aggravate information leakage about $\B$'s documents.

We now show yet another attack that lets $\A$ learn even more, since the N-grams representation \emph{embeds} document's 
structure. From the global space of tri-grams ${\cal GS}$, we can construct a directed graph 
$G(V,E)$ representing relations between tri-grams in $\B$'s document collection.  Any path 
in such a graph will lead to a textual fragment contained in some document held by $\B$.
A vertex in the graph represents a tri-gram; whereas, an edge between two vertices implies that 
the two corresponding tri-grams are consecutive tri-grams in a word.
Given a trigram $x\in {\cal GS}$, with $x^{(i)}$ we denote the $i$-th letter in $x$.
The directed graph $G(V,E)$ is constructed as follows. The vertex set is
$V = \{ V_x \mid x\in {\cal GS}\}$ and the edge set is
$E = \{ \langle V_x, V_y\rangle \mid x^{(2)}=y^{(1)} \wedge\ x^{(3)}=y^{(2)} \}$.
A path $V_{x_1},\ldots,V_{x_n}$ in $G$, will correspond to the string 
$x_1^{(1)}x_1^{(2)}x_2^{(3)}x_3^{(3)}\cdots x_n^{(3)}$. Such a string (or some of its
substring) appears in some document in $\B$'s collection.
By using algorithms based on Deep First Search visit of a graph, a vocabulary, and syntactic rules, 
we could extract large document's chunks.
We did not explore further other techniques to extract ``information'' from the global space of 
tri-grams as we consider them to be out of the scope of this paper. %

\end{document}